\documentclass[aps,12pt,showpacs,preprint,groupedaddress,floatfix]{revtex4}
\usepackage{graphicx}
\usepackage{dcolumn}
\usepackage{bm}
\usepackage{amssymb}
\usepackage{amsmath}
\usepackage{multirow}

\begin{document}

\title{Beyond the relativistic mean-field approximation (III): collective Hamiltonian
 in five dimensions}
\author{T. Nik\v si\' c}
\author{Z. P. Li}
\altaffiliation{School of Physics, Peking University, Beijing, China}
\author{D. Vretenar}
\affiliation{Physics Department, Faculty of Science, University of Zagreb,
Croatia}
\author{L. Pr\'{o}chniak}
\affiliation{Institute of Physics, Maria Curie-Sklodowska University, Lublin, Poland}
\author{J. Meng}
\affiliation{School of Physics, Peking University, Beijing, China}
\author{P. Ring}
\affiliation{Physik-Department der Technischen Universit\"at M\"unchen,
Garching, Germany}
\date{\today}

\begin{abstract}
The framework of relativistic energy density functionals is extended to include
correlations related to restoration of broken symmetries and fluctuations
of collective variables. A new implementation is developed for the solution of the
eigenvalue problem of a five-dimensional collective Hamiltonian for
quadrupole vibrational and rotational degrees of freedom, with parameters
determined by constrained self-consistent relativistic mean-field calculations
for triaxial shapes. The model is tested in a series of illustrative calculations
of potential energy surfaces and the resulting collective excitation spectra
and transition probabilities of the chain of even-even gadolinium isotopes.
\end{abstract}

\maketitle

\section{\label{secI}Introduction}
Nuclear structure models based on energy density functionals (EDFs)
have successfully been used over the whole nuclide chart, from
relatively light systems to superheavy nuclei, and from the valley of
$\beta$-stability to the particle drip-lines \cite{BHR.03,VALR.05,Meng.06}.
In lowest order -- the mean-field approximation, an EDF is
constructed as a functional of one-body nucleon density matrices that
correspond to a single product state -- Slater determinant of
single-particle or single-quasiparticle states. This framework can
thus also be referred to as single reference (SR) EDF. The static
nuclear mean-field is characterized by symmetry breaking --
translational, rotational, particle number. Even though symmetry
breaking incorporates important static correlations, i.e.
deformations and pairing in the SR EDF, this framework can only
describe ground-state properties: binding energies, charge radii,
etc. Excitation spectra and electromagnetic transition probabilities
can only be calculated by including correlations beyond the static
mean-field through restoration of broken symmetries and configuration
mixing of symmetry-breaking product states. The most effective
approach to configuration mixing calculations is the generator
coordinate method (GCM)~\cite{RS.80}, with multipole moments used as
collective coordinates that generate the symmetry-breaking product
wave functions. In such a multi-reference (MR) EDF approach, families
of static mean-field configurations are mixed to restore symmetries
and take into account fluctuations of collective variables. The
corresponding EDFs are functionals of transition densities built from
pairs of symmetry-breaking product states.

In the first two parts of this work \cite{CM_I,CM_II} we have
extended the framework of relativistic energy density
functionals to include correlations related to the restoration of
broken symmetries and to fluctuations of collective variables. A
model has been developed that uses the GCM to perform configuration
mixing of angular-momentum and particle-number projected relativistic
wave functions. The geometry is restricted to axially symmetric
shapes, and the intrinsic wave functions are generated from the
solutions of the relativistic mean-field + Lipkin-Nogami BCS
equations, with a constraint on the mass quadrupole moment. The model
employs a relativistic point-coupling (contact) nucleon-nucleon
effective interaction in the particle-hole channel, and a
density-independent $\delta$-interaction in the particle-particle
channel. This approach enables a quantitative description of the
evolution of shell-structure, deformation and shape coexistence
phenomena in nuclei with soft potential energy surfaces.

In the first application~\cite{PRL99}, the GCM based on relativistic
EDFs was employed in a study of shape transitions in Nd
isotopes. It has been shown that the microscopic framework based on
universal EDFs, adjusted to nuclear ground-state properties, and
extended to take into account correlations related to symmetry
restoration and fluctuations of collective variables, describes not
only general features of shape transitions, but also the unique
behavior of the excitation spectra and transition rates at the X(5)
critical point of quantum shape phase transition. However, an exact 
diagonalization of the X(5) Hamiltonian carried out in Ref.~\cite{caprio}, 
has shown that many properties of the solution are dominated by 
$\beta - \gamma$ coupling induced by the kinetic energy operator.
The importance of the explicit treatment of the triaxial degree of 
freedom, i.e. inclusion of $\beta - \gamma$ coupling, was also 
emphasized in two recent studies of shape transitions in the 
Nd isotopic chain \cite{RE.08,RRS.08}, that used the self-consistent 
Hartree-Fock-Bogoliubov model, based on the finite-range and 
density-dependent Gogny interaction, to generate potential energy 
surfaces in the $\beta - \gamma$ plane.

While GCM configuration mixing of axially symmetric mean-field states
has been implemented by several groups and routinely used in nuclear
structure studies, the application of this method to triaxial shapes
is a much more difficult problem. Only very recently a model has been
introduced~\cite{BH.08}, based on the mean-field states generated by
triaxial quadrupole constraints that are projected on particle number
and angular momentum and mixed by the generator coordinate method.
This method is equivalent to a seven-dimensional GCM calculation,
mixing all five degrees of freedom of the quadrupole operator and the
gauge angles for protons and neutrons. However, the numerical
implementation of the model is very complex, and applications to
medium-heavy and heavy nuclei are still computationally too demanding
and time consuming. In addition, the use of general EDFs, i.e. with
an arbitrary dependence on nucleon densities, in GCM type
calculations, often leads to discontinuities or even divergences of
the energy kernels as a function of deformation~\cite{AER.01,Dob.07}.
Only for certain types of density dependence a regularization method
can be implemented~\cite{LDB.08}, which corrects energy kernels and
removes the discontinuities and divergences. 

In an alternative approach to five-dimensional quadrupole dynamics
that includes rotational symmetry restoration and takes into account
triaxial quadrupole fluctuations, a collective Bohr Hamiltonian is
constructed, with deformation-dependent parameters determined from
microscopic self-consistent mean-field calculations
~\cite{RG.87,Bon.90}. The collective Hamiltonian can be derived in
the Gaussian overlap approximation (GOA)~\cite{RS.80} to the full
five-dimensional GCM. With the assumption that the GCM overlap kernels can be
approximated by Gaussian functions, the local expansion of the
kernels up to second order in the non-locality transforms the
GCM Hill-Wheeler equation into a second-order differential equation -
the Schr\"odinger equation for the collective Hamiltonian. The
kinetic part of this Hamiltonian contains an inertia 
tensor \cite{GG.75}, and the potential energy is determined by the 
diagonal elements of the Hamiltonian kernel, and also includes zero-point 
energy (ZPE) corrections \cite{GG.79}. The adiabatic time-dependent 
Hartree-Fock (ATDHF) theory \cite{BV.78} provides an alternative way to derive 
a classical collective Hamiltonian and, after re-quantization, a
Bohr Hamiltonian of the same structure is obtained, but with different
microscopic expressions for the inertia parameters \cite{Vil.77}.
There is a long-standing debate in the literature about masses in the 
collective Bohr Hamiltonian \cite{GR.80}, i.e. whether the GCM-GOA
expressions (the so-called Yoccoz masses~\cite{PY.57}) 
or the ATDHF expressions (the so-called Thouless-Valatin 
masses~\cite{TV.62}) should be used. The Thouless-Valatin masses
have the advantage that they also include the time-odd components of
the microscopic wave functions and, in this sense, the full dynamics of a
nuclear system. In the GCM approach these components can only be included if,
in addition to the coordinates  $q_i$, the corresponding
canonically conjugate momenta $p_i$ are also taken into account, but this 
is obviously a very complicated task. In many applications a further
simplification is thus introduced in terms of cranking formulas \cite{Ing.56,GG.79},
i.e. the perturbative limit for the Thouless-Valatin masses, and the
corresponding expressions for ZPE corrections. This approximation was 
applied in recent studies using models based both on the Gogny
interaction \cite{LGD.99}, and Skyrme energy density functionals \cite{Pro.04}.

In this work we develop a new implementation for the
solution of a five-dimensional collective Hamiltonian that describes
quadrupole vibrational and rotational degrees of freedom, with
parameters determined in the framework of relativistic EDF. An initial
study along this line which, however, did not include ZPE corrections, 
was reported in Ref.~\cite{PR.04}.

The theoretical framework is described in
Sec.~\ref{secII}: the method of solution of the eigenvalue problem of
the general collective Hamiltonian, and the calculation of the mass
parameters, moments of inertia, and ZPE corrections. In
Sec.~\ref{secIII} the model is tested in the calculation of
collective excitation spectra of the chain of even-even Gd isotopes,
and results are compared with available data. Sec.~\ref{secIV}
presents a summary and an outlook for future studies. Technical
details about the solution of the Dirac equation in triaxial
geometry, the calculation of moments of inertia, ZPE
corrections, and numerical tests are included in Appendix A-D.

\section{\label{secII} Theoretical framework}
\subsection{\label{secIIa}Collective Hamiltonian
 in five dimensions}

Nuclear excitations determined by quadrupole vibrational and
rotational degrees of freedom can be treated simultaneously by
considering five quadrupole collective coordinates 
$\alpha_\mu,\;\mu =-2,-1,\dots,2$ that describe the surface of
a deformed nucleus: $R=R_0[1+\sum_{\mu}\alpha_\mu Y^*_{2\mu}]$. 
To separate rotational and vibrational motion,
these coordinates are usually parameterized in terms of two
deformation parameters $\beta$ and $\gamma$, and three Euler angles
$(\phi,\;\theta,\;\psi)\equiv \Omega$ which define the orientation
of the intrinsic principal axes in the laboratory frame
\begin{equation}
\alpha_\mu = D_{\mu 0}^2(\Omega)\beta\cos{\gamma}+\frac{1}{\sqrt{2}}
   \left[D_{\mu  2}^2(\Omega)+D_{\mu -2}^2(\Omega)\right]\beta \sin{\gamma}\; ,
\end{equation}
where $D^\lambda_{\mu \nu}$ is the Wigner function~\cite{Var.88}. The
three terms of the classical collective Hamiltonian, expressed in
terms of the intrinsic variables $\beta$, $\gamma$ and Euler angles
\begin{equation}
\label{hamiltonian-cl}
H_{\textnormal{coll}} = \mathcal{T}_{\textnormal{vib}}(\beta,\gamma)
                            +\mathcal{T}_{\textnormal{rot}}(\beta,\gamma,\Omega)
                            +\mathcal{V}_{\textnormal{coll}}(\beta,\gamma)\; ,
\end{equation}
denote the contributions from the vibrational kinetic energy:
\begin{equation}
\mathcal{T}_{\textnormal{vib}} = \frac{1}{2}B_{\beta\beta}\dot{\beta}^2
   + \beta B_{\beta\gamma} \dot{\beta}\dot{\gamma}
   +\frac{1}{2} \beta^2B_{\gamma\gamma}\dot{\gamma}^2\; ,
\end{equation}
the rotational kinetic energy:
\begin{equation}
\mathcal{T}_{\textnormal{rot}} = \frac{1}{2}\sum_{k=1}^3{\mathcal{I}_k\omega_k^2},
\end{equation}
and the collective potential energy
$\mathcal{V}_{\textnormal{coll}}(\beta,\gamma)$. The mass parameters
$B_{\beta\beta}$, $B_{\beta\gamma}$, $B_{\gamma\gamma}$, and the
moments of inertia $\mathcal{I}_k$ depend on the quadrupole
deformation variables $\beta$ and $\gamma$.

The Hamiltonian Eq.~(\ref{hamiltonian-cl}) is quantized according to
the general Pauli prescription~\cite{Pauli.33}: for the classical
kinetic energy
\begin{equation}
T=\frac{1}{2}\sum_{ij}{B_{ij}(q)\dot{q}_i\dot{q}_j} \; ,
\end{equation}
and the corresponding quantized form reads:
\begin{equation}
\hat{H}_{\textnormal{kin}} = -\frac{\hbar^2}{2}\frac{1}{\sqrt{\textnormal{det} B}}
                          \sum_{ij}{\frac{\partial}{\partial q_i}
         \sqrt{\textnormal{det} B}(B^{-1})_{ij}\frac{\partial}{\partial q_j} }.
\end{equation}
The kinetic energy tensor in Eq. (\ref{hamiltonian-cl}) takes the block diagonal form:
\begin{equation}
B = \left( \begin{array}{cc} B_{\textnormal{vib}} & 0 \\ 
                                                  0 & B_{\textnormal{rot}} \end{array} \right) \;,
\end{equation}
with the vibrational part of the tensor
\begin{equation}
B_{\textnormal{vib}} = \left( \begin{array}{cc} B_{\beta\beta} & \beta B_{\beta\gamma} \\
                                          \beta B_{\beta\gamma} &\beta^2 B_{\gamma \gamma}
        \end{array} \right).
\end{equation}
In general the rotational part is a complicated function of the Euler
angles but, using the quasi-coordinates related to the components of
the angular momentum in the body-fixed frame, it takes a simple
diagonal form
\begin{equation}
\left(B_{\textnormal{rot}}\right)_{ik}=\delta_{ik}\mathcal{I}_k, \quad k=1,2,3 \;,
\end{equation}
with the moments of inertia expressed as
\begin{equation}
\mathcal{I}_k = 4B_k\beta^2\sin^2(\gamma-2k\pi/3) \;.
\end{equation}
This particular functional form is motivated by the fact that
all three moments of inertia vanish for the spherical configuration
($\beta=0$) and, additionally, $\mathcal{I}_z$ and $\mathcal{I}_y$ vanish for axially
symmetric prolate ($\gamma=0^0$) and oblate ($\gamma=60^0$)
configurations, respectively.
The  resulting determinant reads
\begin{equation}
\label{detB}
\textnormal{det} B = \textnormal{det} B_{\textnormal{vib}}\cdot 
                                \textnormal{det} B_{\textnormal{rot}}
  = 4 wr\beta^8 \sin^2{3\gamma} \; ,
\end{equation}
where $w=B_{\beta\beta}B_{\gamma\gamma}-B_{\beta\gamma}^2 $ and
   $r=B_1B_2B_3$.
The quantized collective Hamiltonian can hence be written in the form:
\begin{equation}
\label{hamiltonian-quant}
\hat{H} = \hat{T}_{\textnormal{vib}}+\hat{T}_{\textnormal{rot}}
              +V_{\textnormal{coll}} \; ,
\end{equation}
with
\begin{align}
\hat{T}_{\textnormal{vib}} =&-\frac{\hbar^2}{2\sqrt{wr}}
   \left\{\frac{1}{\beta^4}
   \left[\frac{\partial}{\partial\beta}\sqrt{\frac{r}{w}}\beta^4
   B_{\gamma\gamma} \frac{\partial}{\partial\beta}
   - \frac{\partial}{\partial\beta}\sqrt{\frac{r}{w}}\beta^3
   B_{\beta\gamma}\frac{\partial}{\partial\gamma}
   \right]\right.
   \nonumber \\
   &+\frac{1}{\beta\sin{3\gamma}}\left.\left[
   -\frac{\partial}{\partial\gamma} \sqrt{\frac{r}{w}}\sin{3\gamma}
      B_{\beta \gamma}\frac{\partial}{\partial\beta}
    +\frac{1}{\beta}\frac{\partial}{\partial\gamma} \sqrt{\frac{r}{w}}\sin{3\gamma}
      B_{\beta \beta}\frac{\partial}{\partial\gamma}
   \right]\right\} \; ,
\end{align}
and
\begin{equation}
\hat{T}_{\textnormal{\textnormal{\textnormal{rot}}}} = 
\frac{1}{2}\sum_{k=1}^3{\frac{\hat{J}^2_k}{\mathcal{I}_k}} \; ,
\end{equation}
where $\hat{J}_k$ denotes the components of the angular momentum in
the body-fixed frame of a nucleus. $V_{\textnormal{coll}}$ is the collective
potential. The Hamiltonian describes quadrupole vibrations,
rotations, and the coupling of these collective modes. The
determinant Eq.~(\ref{detB}) determines the volume element in the
collective space:
\begin{equation}
\label{measure}
\int{d\tau_{coll}}=\int{d\Omega d\tau_0\sqrt{wr}}=
\int_0^\infty{d\beta \beta^4 \int_0^{2\pi}{d\gamma|\sin{3\gamma}|
         \int{d\Omega \sqrt{wr}}}} \; ,
\end{equation}
and the quantized Hamiltonian Eq.~(\ref{hamiltonian-quant}) is
hermitian with respect to the collective measure Eq.~(\ref{measure}).

The methods used to solve the eigenvalue problem of the general
collective Hamiltonian Eq.~(\ref{hamiltonian-quant}) can be divided
into two classes. The first is based on a direct numerical solution
of a system of partial differential equations using finite-difference
methods~\cite{KB.67,Roh.77,Tro.92}. The second approach uses an
expansion of eigenfunctions in terms of a complete set of basis
functions, that depend on the deformation variables $\beta$ and
$\gamma$, and the Euler angles $\phi$, $\theta$ and $\psi$
~\cite{DB.70,GG.71,Kum.74,LQ.82}. The eigenvalue problem reduces
to a simple matrix diagonalization, and the main task is the
construction of an appropriate basis for each value of the angular
momentum quantum number.

In this work we employ the second approach and construct basis states
according to the method described in Refs.
~\cite{LQ.82,Del.89,DLB.89,Pro.99,LGD.99}. For each value of the
angular momentum $I$, one chooses a complete set of square integrable
functions
\begin{equation}
\label{basis-start}
\phi_{Lmn}^{IM}(\beta,\gamma,\Omega)=e^{-\mu^2\beta^2/2}\beta^n
 \left\{ \begin{array}{c} \cos{m\gamma} \\ \sin{m\gamma}\end{array}\right\}
 {D_{ML}^{I*}}(\Omega).
\end{equation}
The projections $M$ and $L$ are determined by the angular momentum:
$M,L=-I,\dots,I$. In principle, the parameter $n$ can take any
nonnegative integer value, but in actual calculations a certain
cut-off value $n_{max}$ has to be imposed. The allowed values of $m$
are: $m=n,n-2,\dots,0$ or $1$. The choice of the function
$e^{-\mu^2\beta^2/2}$ ensures that the basis states generate wave
functions that vanish at large deformations ($\beta \to \infty$). The
basis parameter $\mu$ has to be adjusted for each nucleus
individually, so that it minimizes the ground state energy of the
nucleus. However,  if the cut-off value $n_{max}$ is large enough, a
stable ground-state solution can be found for a broad range of values
of the parameter $\mu$ .

The basis states have to fulfill certain symmetry conditions that
originate from the fact that the choice of the body-fixed frame is
not unique. For a given quadrupole tensor  $\alpha_\mu$ in the
laboratory frame, there are 24 possible orientations of the
body-fixed right-hand coordinate system, corresponding to different
values of the variables $\beta$, $\gamma$, and $\Omega$. The basis
states in the body-fixed frame must be invariant with respect to the
transformations that connect various choices of the body-fixed frame,
and which form a finite group isomorphic to the octahedral point
group $O$~\cite{KB.67,Ham.62} (group of proper rotations which take a
cube or octahedron into itself). This symmetry condition is fulfilled
by linear combinations of the states (\ref{basis-start})
\begin{equation}
\label{basis-oct}
\xi_{Lmn}^{IM}(\beta,\gamma,\Omega)=e^{-\mu^2\beta^2/2}\beta^n
  \sum_{K \in \Delta I}{f_{LmK}^I(\gamma)\Phi_{MK}^I(\Omega)} \; ,
\end{equation}
invariant under the transformations of the octahedral group. The
angular part corresponds to linear combinations of the Wigner
functions
\begin{equation}
\label{Wigner}
\Phi_{MK}^I(\Omega)=\sqrt{\frac{2I+1}{16\pi^2(1+\delta_{K0})}}
\left[D_{MK}^{I*}(\Omega)+(-1)^ID_{M-K}^{I*}(\Omega) \right] \; ,
\end{equation}
and the summation in Eq. (\ref{basis-oct}) is over the allowed set  of
the $K$ values:
\begin{equation}
\Delta I = \left\{ \begin{array}{c}
   0,2,\dots,I \quad \textnormal{for} \quad  I\; \textnormal{mod}\; 2 = 0 \\
   2,4,\dots,I-1 \quad \textnormal{for} \quad   I\; \textnormal{mod}\; 2 =1
\end{array} \right.
\end{equation}
In the next step linearly independent functions have to be selected
from the overcomplete basis set Eq.~(\ref{basis-oct}). In addition,
some of the basis states have to be discarded in order to enforce the
correct behavior of solutions on the $\gamma=n\pi/3$
axes~\cite{KB.67}. A simple and elegant solution of both problems is
provided by group theoretical methods~\cite{Pro.XX}.  Finally, the
basis states Eq.~(\ref{basis-oct}) are not orthogonal. Although the
Hamiltonian could also be diagonalized directly in a non-orthogonal
basis~\cite{Kum.74}, we choose to orthogonalize the basis states by
applying the Cholesky-Banachiewicz method~\cite{NR}.

The diagonalization of the collective Hamiltonian yields the 
energy spectrum $E_\alpha^I$ and the corresponding eigenfunctions
\begin{equation}
\label{wave-coll}
\Psi_\alpha^{IM}(\beta,\gamma,\Omega) =
  \sum_{K\in \Delta I}
           {\psi_{\alpha K}^I(\beta,\gamma)\Phi_{MK}^I(\Omega)}.
\end{equation}
Using the collective wave functions Eq.~(\ref{wave-coll}), 
various observables can be calculated and compared with
experimental results. For instance, the quadrupole E2 reduced
transition probability:
\begin{equation}
\label{BE2}
B(\textnormal{E2};\; \alpha I \to \alpha^\prime I^\prime)=
      \frac{1}{2I+1}|\langle \alpha^\prime I^\prime || \mathcal{\hat{M}}(E2) ||
                                        \alpha I  \rangle|^2 \; ,
\end{equation}
and the spectroscopic quadrupole moment of the state $|\alpha I\rangle$:\begin{equation}
\label{Qspec}
Q_{\textnormal{spec},\alpha I} = \frac{1}{\sqrt{2I+1}}C_{II20}^{II}
     \langle \alpha I || \mathcal{\hat{M}}(E2) ||\alpha I  \rangle \; ,
\end{equation}
where $\mathcal{\hat{M}}(E2)$ denotes the electric quadrupole operator.
Detailed expressions for the reduced matrix element
$\langle \alpha^\prime I^\prime || \mathcal{\hat{M}}(E2) || \alpha I  \rangle $
can be found  in Ref.~\cite{KB.67}.

The shape of a nucleus can be characterized in a qualitative way
by average values of the
invariants $\beta^2$, $\beta^3\cos{3\gamma}$, as well as their
combinations. For example, the average value of the invariant
$\beta^2$ in the state $|\alpha I\rangle$:
\begin{equation}
\langle \beta^2\rangle_{I\alpha} = \langle \Psi_\alpha^I | \beta^2 | \Psi_\alpha^I\rangle
=\sum_{K\in\Delta I}{\int{\beta^2|\psi^I_{\alpha,K}(\beta,\gamma)|^2d\tau_0}} \; ,
\end{equation}
and the average values of the deformation parameters
$\beta$ and $\gamma$ in the state
$|\alpha I\rangle$ are calculated from:
\begin{align}
\label{avbeta}
\langle \beta\rangle_{I\alpha} &= \sqrt{\langle \beta^2\rangle_{I\alpha} }, \\
\label{avgamma}
\langle \gamma\rangle_{I\alpha} &=
        \frac{1}{3}\arccos{\frac{\langle \beta^3 \cos{3\gamma}\rangle_{I\alpha}}
            {\sqrt{\langle \beta^2\rangle_{I\alpha} \langle \beta^4\rangle_{I\alpha}}}} ; .
\end{align}
The mixing of different intrinsic configurations in the state
$|\alpha I\rangle$ can be determined from
the distribution of the projection $K$ of the angular momentum $I$
on the $z$ axis in the body-fixed frame:
\begin{equation}
N_K=6\int_0^{\pi/3}{\int_0^\infty{
   |\psi^I_{\alpha,K}(\beta,\gamma)|^2\beta^4|\sin{3\gamma}|d\beta d\gamma}},
\label{NK}
\end{equation}
where the components $\psi^I_{\alpha,K}(\beta,\gamma)$ are defined in 
Eq.~(\ref{wave-coll}).
For large deformations the $K$ quantum number is to a good approximation
conserved. Consequently, only one of the integrals Eq.~(\ref{NK})
will give a value close to $1$. A broader distribution of $N_K$
values in the state $|\alpha I\rangle$ provides a measure of mixing
of intrinsic configurations.

\subsection{\label{secIIb}Parameters of the collective Hamiltonian}

The entire dynamics of the collective Hamiltonian is governed by the
seven functions of the intrinsic deformations $\beta$ and $\gamma$:
the collective potential, the three mass parameters:
$B_{\beta\beta}$, $B_{\beta\gamma}$, $B_{\gamma\gamma}$, and the
three moments of inertia $\mathcal{I}_k$. These functions are
determined by the choice of a particular microscopic nuclear energy
density functional or effective interaction. As in our previous two
studies of configuration mixing effects~\cite{CM_I,CM_II}, also in
this work we use the relativistic functional PC-F1 (point-coupling
Lagrangian)~\cite{BMM.02} in the particle-hole channel, and a
density-independent $\delta$-force is the effective interaction in
the particle-particle channel. The parameters of the PC-F1 functional
and the pairing strength constants $V_n$ and $V_p$ have been adjusted
simultaneously to the nuclear matter equation of state, and to
ground-state observables (binding energies, charge and diffraction
radii, surface thickness and pairing gaps) of spherical nuclei
~\cite{BMM.02}, with pairing correlations treated in the BCS
approximation.

The choice of the point-coupling effective Lagrangian determines the
self-consistent relativistic mean-field energy (RMF) of a nuclear
system in terms of local single-nucleon densities and currents:
\begin{align}
{{E}}_{\textnormal{RMF}} &=
 \int d{\bm r }~{\mathcal{E}_{\textnormal{RMF}}}(\bm{r}) \nonumber \\
          &=\sum_k{\int d\bm{r}~v_k^2~{\bar{\psi}_k (\bm{r}) \left( -i\bm{\gamma}
      \bm{\nabla} + m\right )\psi_k(\bm{r})}} \nonumber \\
 &+ \int d{\bm r }~{\left (\frac{\alpha_S}{2}\rho_S^2+\frac{\beta_S}{3}\rho_S^3 +
  \frac{\gamma_S}{4}\rho_S^4+\frac{\delta_S}{2}\rho_S\triangle \rho_S
 + \frac{\alpha_V}{2}j_\mu j^\mu + \frac{\gamma_V}{4}(j_\mu j^\mu)^2 +
       \frac{\delta_V}{2}j_\mu\triangle j^\mu \right.} \nonumber \\
 &+ \left .
  \frac{\alpha_{TV}}{2}j^{\mu}_{TV}(j_{TV})_\mu+\frac{\delta_{TV}}{2}
    j^\mu_{TV}\triangle  (j_{TV})_{\mu}
 + \frac{\alpha_{TS}}{2}\rho_{TS}^2+\frac{\delta_{TS}}{2}\rho_{TS}\triangle
      \rho_{TS} +\frac{e}{2}\rho_p A^0
 \right) \; ,
\label{EMF}
\end{align}
where $\psi$ denotes the Dirac spinor field of a nucleon. The local
isoscalar (S) and isovector scalar (TS) densities, and corresponding
isoscalar and isovector (TV) currents for a nucleus with A nucleons
\begin{align}
\label{dens_1}
\rho_{S}({\bm r}) &=\sum_k v_k^2 ~\bar{\psi}_{k}({\bm r})
             \psi _{k}({\bm r})~,  \\
\label{dens_2}
\rho_{TS}({\bm r}) &=\sum_k v_k^2 ~
      \bar{\psi}_{k}({\bm r})\tau_3\psi _{k}^{{}}({\bm r})~,  \\
\label{dens_3}
j^{\mu}({\bm r}) &=\sum_k v_k^2 ~\bar{\psi}_{k}({\bm r})
        \gamma^\mu\psi _{k}^{{}}({\bm r})~,  \\
\label{dens_4}
j^{\mu}_{TV}({\bm r}) &=\sum_k v_k^2 ~\bar{\psi}_{k}({\bm r})
     \gamma^\mu \tau_3 \psi _{k}^{{}}({\bm r})~,
\end{align}
are calculated in the {\it no-sea} approximation: the summation in
Eqs.~(\ref{EMF}) - (\ref{dens_4}) runs over all occupied states in
the Fermi sea, i.e. only occupied single-nucleon states with positive
energy explicitly contribute to the nucleon densities and currents.
$v_k^2$ denotes the occupation factors of single-nucleon states. In
Eq.~(\ref{EMF}) $\rho_p$ is the proton density, and $A^0$ denotes the
Coulomb potential. $\alpha$, $\beta$, $\gamma$ and $\delta$ denote
the 11 parameters of the PC-F1 relativistic density functional in the
corresponding space-isospace channels.

The single-nucleon wave functions represent self-consistent solutions
of the Dirac equation:
\begin{equation}
\left\{ \bm{\alpha}\cdot\left[ -i\bm{\nabla}-\bm{V}({\bm r}) \right] +V({\bm r})+
\beta \big(m+S({\bm r})\big) \right\} \psi_{i}({\bm r}) = \epsilon_i\psi_{i}({\bm r})\;.
\label{dirac}
\end{equation}
The scalar and vector potentials
\begin{equation}
S({\bm r}) = \Sigma_S({\bm r}) + \tau_3\Sigma_{TS}({\bm r})\;,
\label{scapot}
\end{equation}
\begin{equation}
V^{\mu}({\bm r})  = \Sigma^{\mu}({\bm r}) + \tau_3\Sigma^{\mu}_{TV}({\bm r})\;,
\label{vecpot}
\end{equation}
contain the nucleon isoscalar-scalar, isovector-scalar,
isoscalar-vector and isovector-vector self-energies
defined by the following relations:
\begin{eqnarray}
   \label{selfS}
   \Sigma_S & = & \alpha_S \rho_S + \beta_S \rho_S^2 +
   \gamma_S\rho_S^3+ \delta_S \triangle \rho_S \; ,\\
   \label{selfTS}
   \Sigma_{TS} & = & \alpha_{TS} \rho_{TS}
   +\delta_{TS} \triangle \rho_{TS}\; , \\
   \label{selfV}
   \Sigma^{\mu} & = & \alpha_V j^{\mu}
  +\gamma_V (j_\nu j^\nu)j^\mu + \delta_V \triangle j^\mu
   -eA^\mu\frac{1-\tau_3}{2} \; ,\\
   \label{selfTV}
   \Sigma^{\mu}_{TV} & = & \alpha_{TV} j^{\mu}_{TV}
      + \delta_{TV} \triangle j^{\mu}_{TV}\;,
\end{eqnarray}
respectively. Because of charge conservation, only the $3$-rd
component of the isovector densities and currents contributes to the
nucleon self-energies. In this work we only consider even-even
nuclei, i.e. time-reversal invariance is assumed, which implies
that the spatial components of the single-nucleon currents vanish in
the nuclear ground state.

The Dirac equation (\ref{dirac}) is solved by expanding the nucleon
spinors in the basis of a three-dimensional harmonic oscillator in
Cartesian coordinates. In this way both axial and triaxial nuclear
shapes can be described. In addition, to reduce the computational
task, it is assumed that  the total densities are symmetric under
reflections with respect to all three planes $xy$, $xz$ and $yz$.
When combined with time-reversal invariance, this also implies that
parity is conserved. Under these restrictions we consider only
even-multipole deformations, whereas solutions for odd multipoles
vanish. The method of solution of the Dirac equation is described in
more detail in Appendix~\ref{app-A}.

In addition to the self-consistent mean-field potential, for
open-shell nuclei pairing correlations have to be included in the
energy functional. In this work pairing is treated using the BCS formalism. 
Following the prescription from
Ref.~\cite{BMM.02}, we employ a $\delta$-force in the pairing channel,
supplemented with a smooth cut-off determined by a Fermi function in
the single-particle energies. The pairing contribution to the total
energy is given by
\begin{equation}
{E}_{\textnormal{pair}}^{p(n)} = 
\int{\mathcal{E}_{\textnormal{pair}}^{p(n)}(\bm{r})d\bm{r}}=
   \frac{V_{p(n)}}{4}\int{\kappa_{p(n)}^*(\bm{r})\kappa_{p(n)}(\bm{r}) d\bm{r}}\;,
\label{epair}
\end{equation}
for protons and neutrons, respectively. $\kappa_{p(n)}(\bm{r})$
denotes the local part of the pairing tensor, and $V_{p(n)}$ is the
pairing strength parameter.

The center-of-mass correction is included by adding the expectation
value
\begin{equation}
 E_{cm} = -\frac{\langle \hat{P}_{cm}^2 \rangle}{2mA}\;,
\end{equation}
to the total energy.
Finally, the expression for the total energy reads
\begin{equation}
E_{\textnormal{tot}} =
\int{\left[\mathcal{E}_{\textnormal{RMF}}(\bm{r})
+\mathcal{E}_{\textnormal{pair}}^p(\bm{r})
         +\mathcal{E}_{\textnormal{pair}}^n(\bm{r})\right]d\bm{r} }+E_{cm}\;.
\label{etot}
\end{equation}

The entire map of the energy surface as function of the quadrupole
deformation is obtained by imposing constraints on the axial and
triaxial mass quadrupole moments. The method of quadratic constraints
uses an unrestricted variation of the function
\begin{equation}
\langle H\rangle
   +\sum_{\mu=0,2}{C_{2\mu}\left(\langle \hat{Q}_{2\mu}  \rangle - q_{2\mu}  \right)^2} \; ,
\label{constr}
\end{equation}
where $\langle H\rangle$ is the total energy, and 
$\langle \hat{Q}_{2\mu}\rangle$ denotes the expectation value of the mass quadrupole 
operator:
\begin{equation}
\hat{Q}_{20}=2z^2-x^2-y^2 \quad \textnormal{and}\quad \hat{Q}_{22}=x^2-y^2 \;.
\end{equation}
$q_{2\mu}$ is the constrained value of the multipole moment,
and $C_{2\mu}$ the corresponding stiffness constant~\cite{RS.80}.

The single-nucleon wave functions, energies and occupation factors,
generated from constrained self-consistent solutions of the RMF+BCS
equations, provide the microscopic input for the parameters of the
collective Hamiltonian.

The moments of inertia are calculated according to the Inglis-Belyaev
formula:~\cite{Ing.56,Bel.61}
\begin{equation}
\label{Inglis-Belyaev}
\mathcal{I}_k = \sum_{i,j}{\frac{\left(u_iv_j-v_iu_j \right)^2}{E_i+E_j}
  | \langle i |\hat{J}_k | j  \rangle |^2}\quad k=1,2,3,
\end{equation}
where $k$ denotes the axis of rotation, and the summation runs over
the proton and neutron quasiparticle states. The quasiparticle
energies $E_i$, occupation probabilities $v_i$, and single-nucleon
wave functions $\psi_i$ are determined by solutions of the
constrained RMF+BCS equations. The mass parameters associated with
the two quadrupole collective coordinates
$q_0=\langle\hat{Q}_{20}\rangle$ and $q_2=\langle\hat{Q}_{22}\rangle$
are also calculated in the cranking approximation~\cite{GG.79}
\begin{equation}
\label{masspar-B}
B_{\mu\nu}(q_0,q_2)=\frac{\hbar^2}{2}
 \left[\mathcal{M}_{(1)}^{-1} \mathcal{M}_{(3)} \mathcal{M}_{(1)}^{-1}\right]_{\mu\nu}\;,
\end{equation}
with
\begin{equation}
\label{masspar-M}
\mathcal{M}_{(n),\mu\nu}(q_0,q_2)=\sum_{i,j}
 {\frac{\left\langle i\right|\hat{Q}_{2\mu}\left| j\right\rangle
 \left\langle j\right|\hat{Q}_{2\nu}\left| i\right\rangle}
 {(E_i+E_j)^n}\left(u_i v_j+ v_i u_j \right)^2}\;.
\end{equation}

The collective energy surface includes the energy of the zero-point
motion, which has to be subtracted. The collective zero-point energy
(ZPE) corresponds to a superposition of zero-point motion of
individual nucleons in the single-nucleon potential. In the general
case, the ZPE corrections on the potential energy surfaces depend on
the deformation. The ZPE includes terms originating from the vibrational
and rotational kinetic energy, and a contribution of potential energy
\begin{equation}
\Delta V(q_0,q_2)=\Delta V_{\textnormal{vib}}(q_0,q_2) 
                            + \Delta V_{\textnormal{rot}}(q_0,q_2)
                            + \Delta V_{\textnormal{pot}}(q_0,q_2) \; .
\end{equation}
The latter is much smaller than the contribution of kinetic energy,
and is usually neglected~\cite{LGD.99}. Simple prescriptions for the
calculation of vibrational and rotational ZPE have been derived in
Ref.~\cite{GG.79}. Both corrections are calculated in the cranking
approximation, i.e. on the same level of approximation as the mass
parameters and the moments of inertia. The vibrational ZPE is given
by the expression:
\begin{equation}
\label{ZPE-vib}
\Delta V_{\textnormal{vib}}(q_0,q_2) = \frac{1}{4}
\textnormal{Tr}\left[\mathcal{M}_{(3)}^{-1}\mathcal{M}_{(2)}  \right]\;.
\end{equation}
The rotational ZPE is a sum of three terms:
\begin{equation}
\label{ZPE-rot}
\Delta V_{\textnormal{rot}}(q_0,q_2)=\Delta V_{-2-2}(q_0,q_2)+\Delta V_{-1-1}(q_0,q_2)
                                     +\Delta V_{11}(q_0,q_2),                                    
\end{equation}
with
\begin{equation}
\label{ZPE-rotA}
\Delta V_{\mu\nu}(q_0,q_2) = \frac{1}{4}\frac{\mathcal{M}_{(2),\mu\nu}(q_0,q_2)}
        {\mathcal{M}_{(3),\mu\nu}(q_0,q_2)} \; .
\end{equation}

The individual terms are calculated from Eqs.~(\ref{ZPE-rotA}) and (\ref{masspar-M}),
with the intrinsic components of the quadrupole operator defined by:
\begin{equation}
\hat{Q}_{21}=-2iyz \;,\quad \hat{Q}_{2-1}=-2xz\; ,\quad \hat{Q}_{2-2}=2ixy \; .
\end{equation}
The potential $V_{\textnormal{coll}}$ in the collective Hamiltonian
Eq.~(\ref{hamiltonian-quant}) is obtained by subtracting the ZPE
corrections from the total mean-field energy defined in Eq.~(\ref{etot}):
\begin{equation}
\label{Vcoll}
{V}_{\textnormal{coll}}(q_0,q_2) = E_{\textnormal{tot}}(q_0,q_2)
  - \Delta V_{\textnormal{vib}}(q_0,q_2) - \Delta V_{\textnormal{rot}}(q_0,q_2) \; .
\end{equation}
Detailed expressions for the parameters of the collective Hamiltonian
are given in Appendix~\ref{app-B} and \ref{app-E}.
\section{\label{secIII}Illustrative calculations: the Gadolinium isotopic chain}

\begin{figure}
\includegraphics[scale=0.65,angle=270]{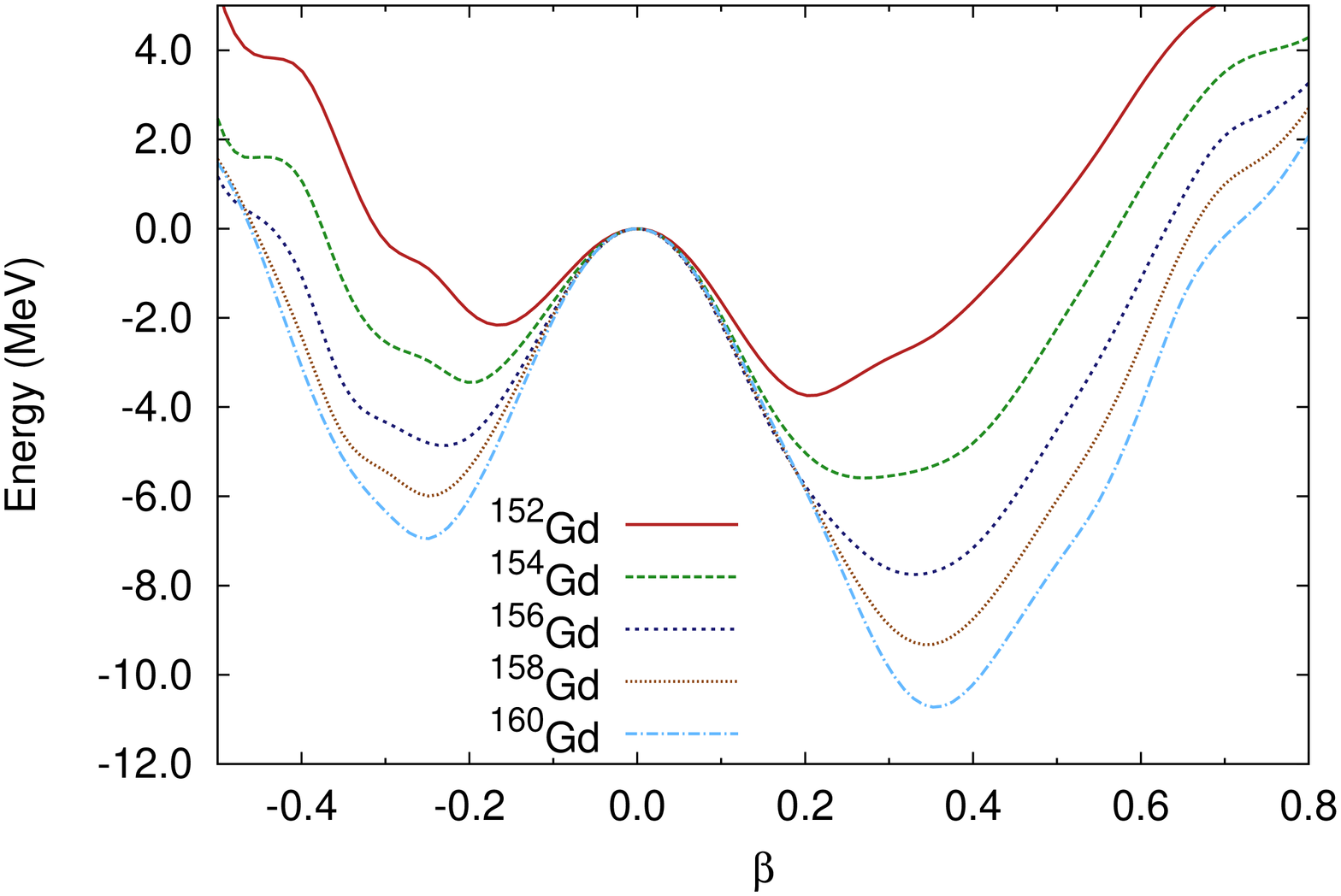}
\vspace{2cm}
\caption{(Color online)
Self-consistent RMF+BCS binding energy curves of the $^{152-160}$Gd
isotopes, as functions of the axial deformation parameter $\beta$. Negative values
of $\beta$ correspond to the $(\beta>0,\;\gamma=180^0)$ axis on the
$\beta - \gamma$ plane.}
\label{Fig1}
\end{figure}

In this section the new implementation is tested in a series of illustrative
calculations of potential energy surfaces and the resulting
collective excitation spectra of the chain of even-even Gd isotopes:
$^{152-160}$Gd. The transition between different shapes, from the
weakly-deformed transitional $^{152}$Gd to the well-deformed prolate
$^{160}$Gd, is illustrated in Fig.~\ref{Fig1}, where we plot the
self-consistent RMF+BCS binding energy curves for the axially
symmetric configurations as functions of the deformation parameter
$\beta$. Negative values of $\beta$ correspond to the
$\beta>0,\;\gamma=180^0$ axis on the $\beta - \gamma$ plane.
$^{152}$Gd is characterized by the coexistence of two weakly-deformed
prolate ($\beta \approx 0.2$) and oblate ($\beta \approx -0.2$)
minima, with the prolate minimum $\approx 4$ MeV below the spherical
configuration. With the addition of more neutrons the deformed minima
become deeper and gradually shift to larger values of $\beta$. For
$^{160}$Gd, the constrained RMF+BCS calculation with the PC-F1
interaction predicts a pronounced prolate minimum at ($\beta \approx
0.35$), more than 10 MeV below the corresponding spherical
configuration.

\begin{figure}
\includegraphics[scale=1.0]{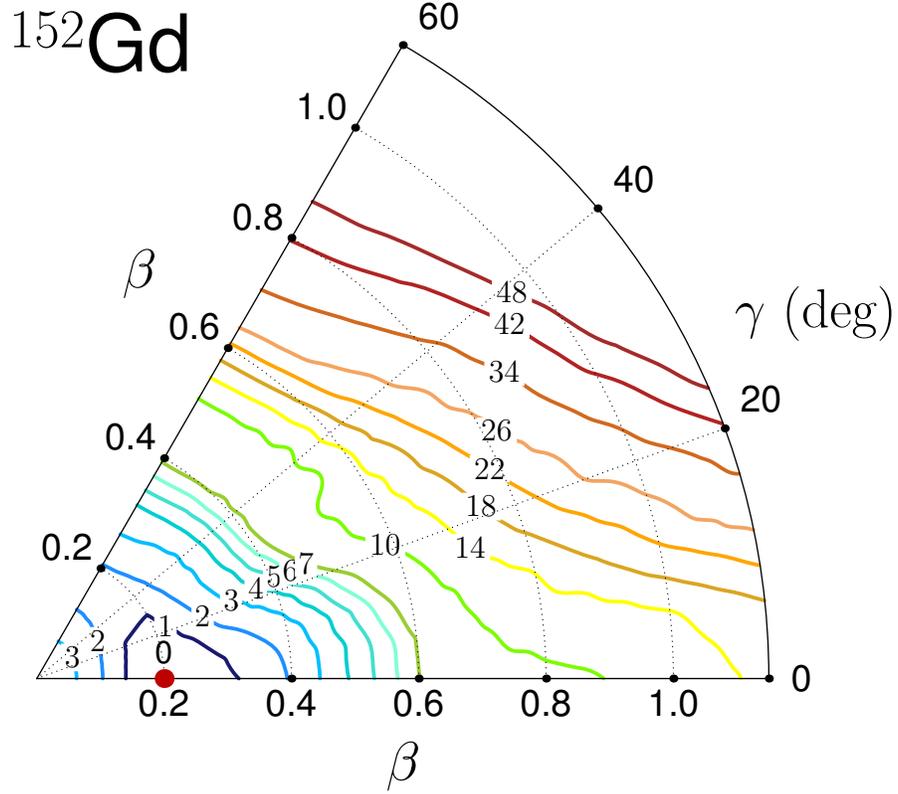}
\caption{(Color online) Self-consistent RMF+BCS triaxial quadrupole
binding energy map of $^{152}$Gd
in the $\beta - \gamma$ plane ($0\le \gamma\le 60^0$).
All energies are normalized with respect to
the binding energy of the absolute minimum (red dot). 
The contours join points
on the surface with the same energy (in MeV).}
\label{Fig2}
\end{figure}

\begin{figure}
\includegraphics[scale=1.0]{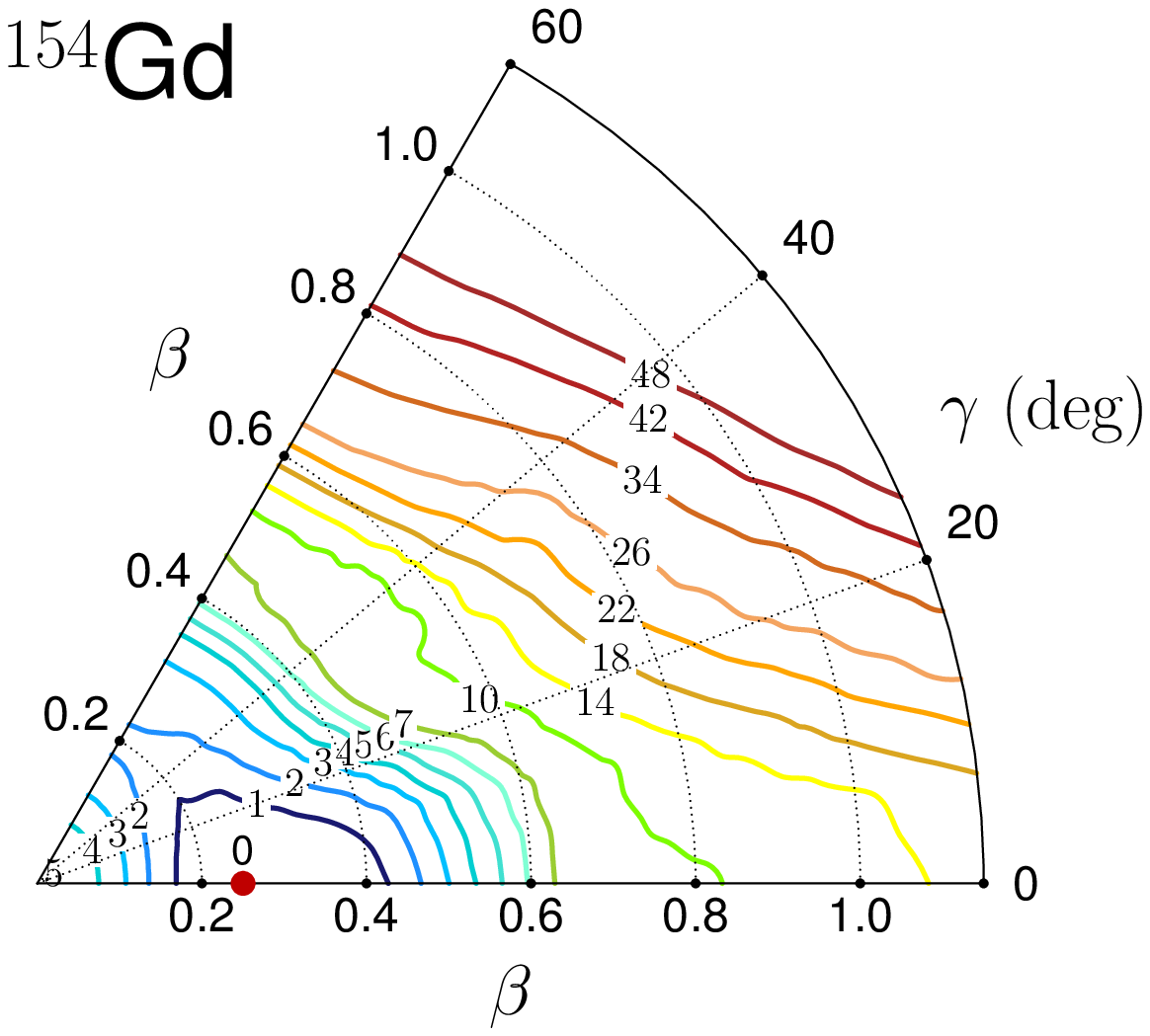}
\caption{(Color online) Same as Fig.~\ref{Fig2}, but for the nucleus $^{154}$Gd.}
\label{Fig3}
\end{figure}

\begin{figure}
\includegraphics[scale=1.0]{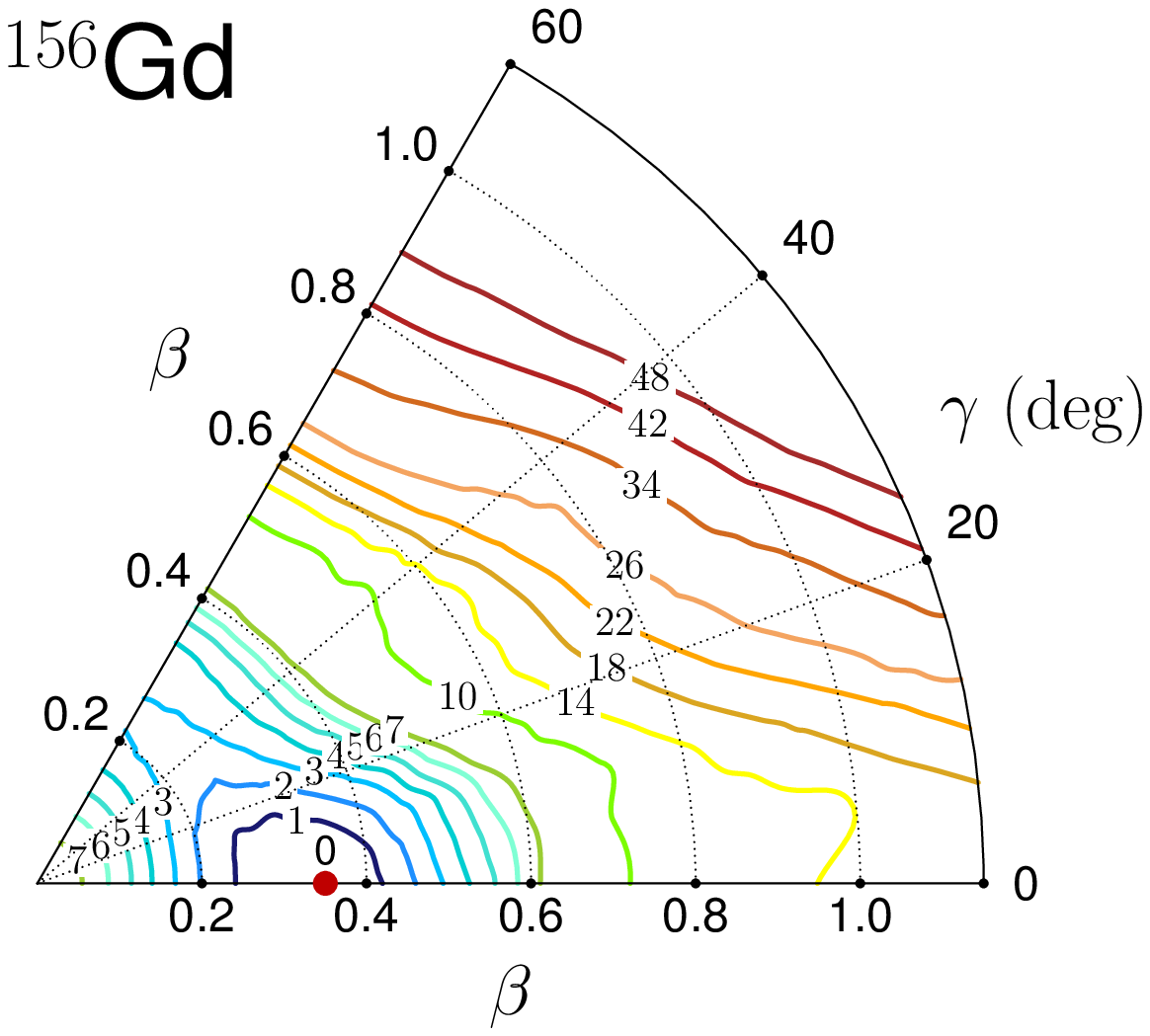}
\caption{(Color online) Same as Fig.~\ref{Fig2}, but for the nucleus $^{156}$Gd.}
\label{Fig4}
\end{figure}

\begin{figure}
\includegraphics[scale=1.0]{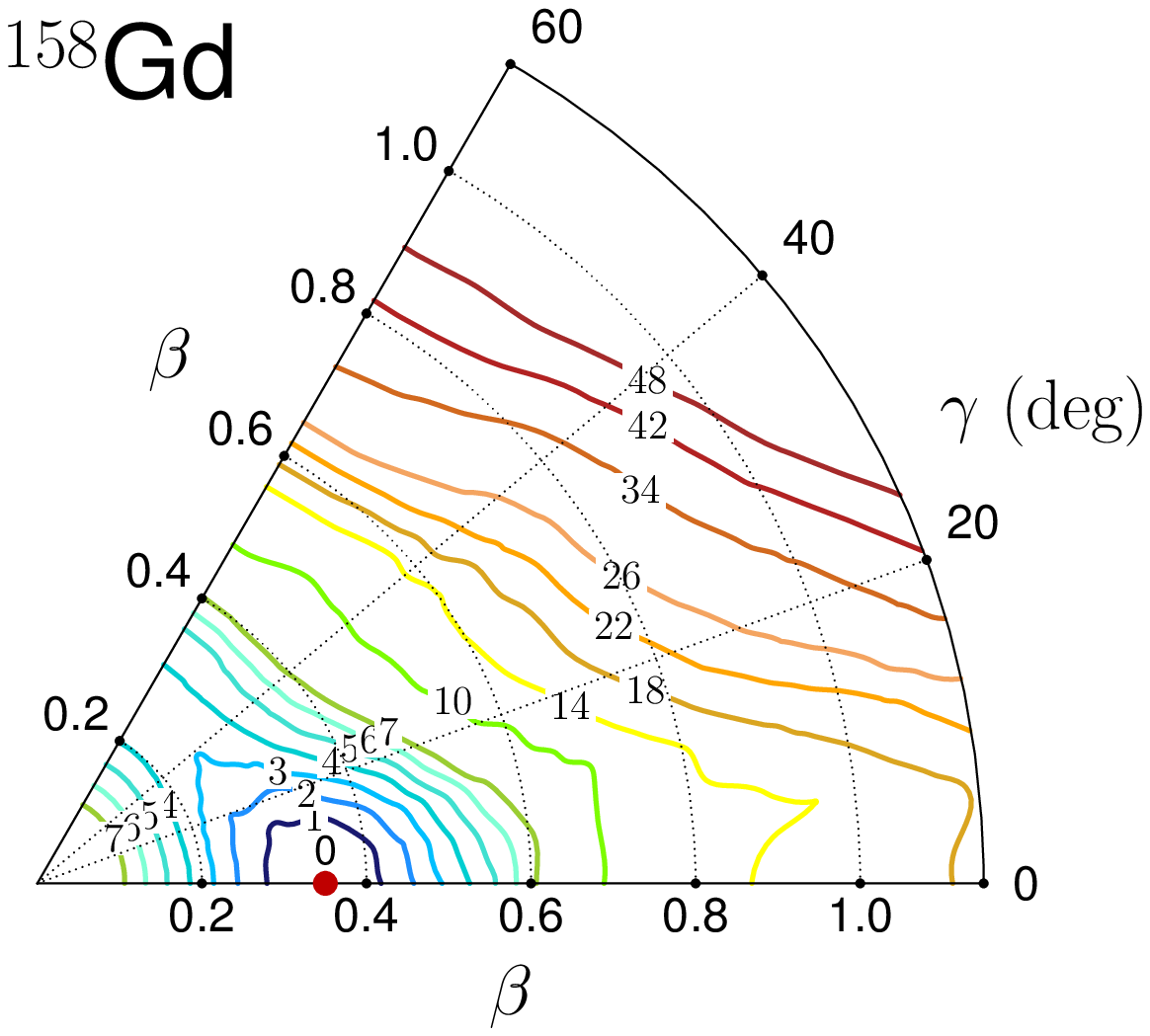}
\caption{(Color online) Same as Fig.~\ref{Fig2}, but for the nucleus $^{158}$Gd.}
\label{Fig5}
\end{figure}

\begin{figure}
\includegraphics[scale=1.0]{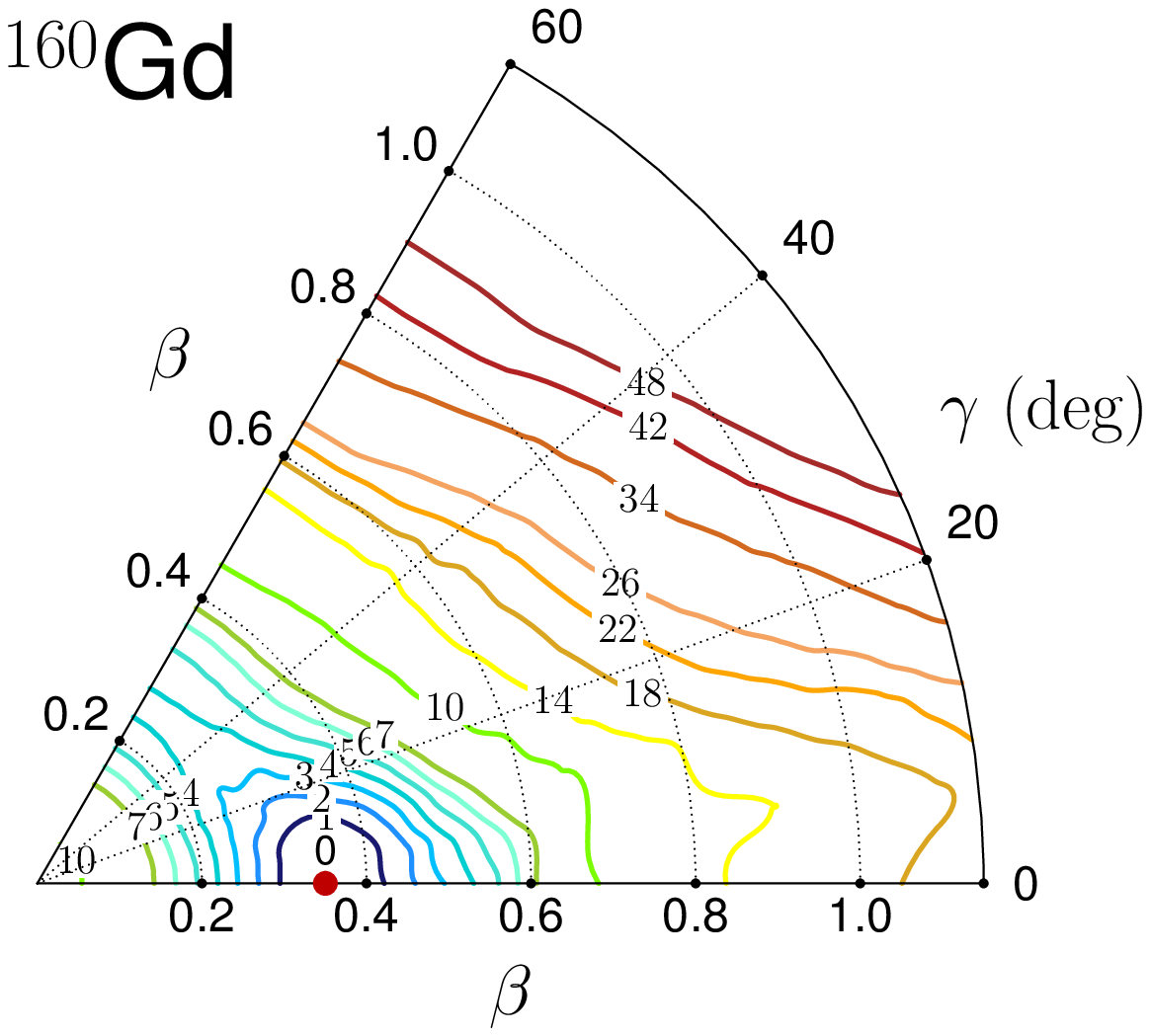}
\caption{(Color online) Same as Fig.~\ref{Fig2}, but for the nucleus $^{160}$Gd.}
\label{Fig6}
\end{figure}

In  Figs.~\ref{Fig2}-\ref{Fig6} we display the self-consistent
RMF+BCS triaxial quadrupole binding energy maps of $^{152-160}$Gd in
the $\beta - \gamma$ plane ($0\le \gamma\le 60^0$), 
obtained by imposing constraints on the expectation
values of the quadrupole moments $\langle \hat{Q}_{2 0}  \rangle$ and
$\langle \hat{Q}_{2 2}  \rangle$ (cf. Eq.~(\ref{constr}) ). Filled circles
denote absolute minima; all energies are normalized with respect to
the binding energy of the absolute minimum. The contours join points
with the same energy. The energy maps nicely illustrate the
gradual increase of deformation of the prolate minimum with increasing
number of neutrons. One notes, however, that the oblate minima
shown in the axial plots in Fig.~\ref{Fig1} are not
true minima, but rather saddle points in the $\beta - \gamma$ plane.

\begin{figure}
\includegraphics[scale=1.1]{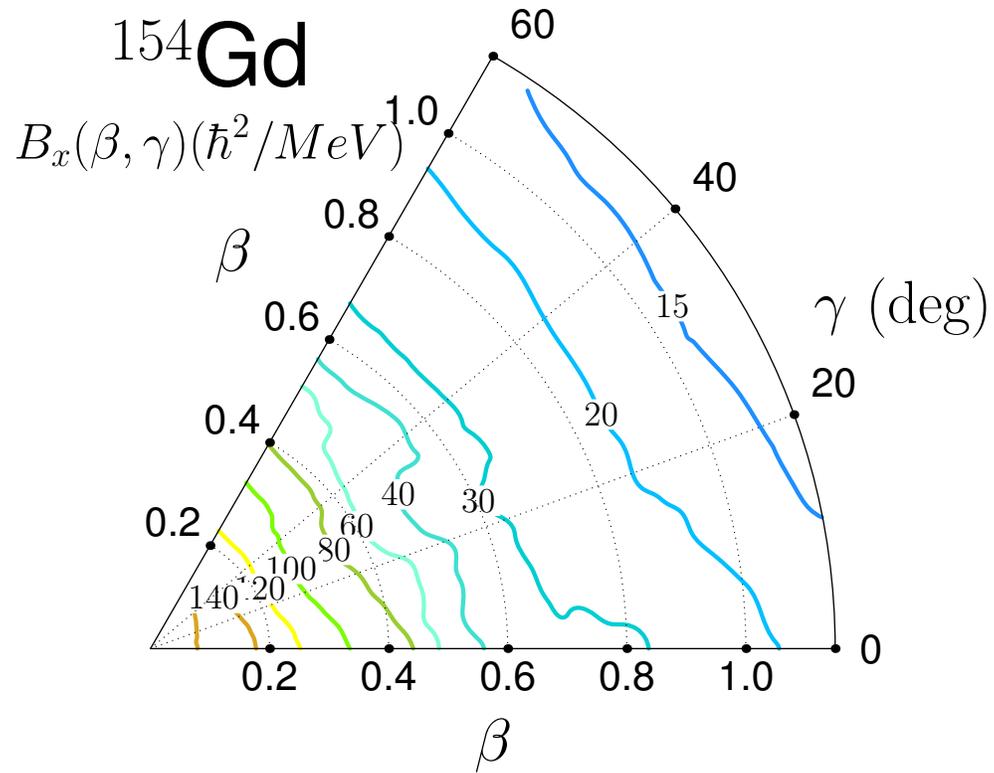}
\caption{(Color online) The map of the Inglis-Belyaev inertia parameter $B_x$
of $^{154}$Gd in the $\beta - \gamma$ plane ($0\le \gamma\le 60^0$).}
\label{Fig7}
\end{figure}

\begin{figure}
\includegraphics[scale=1.1]{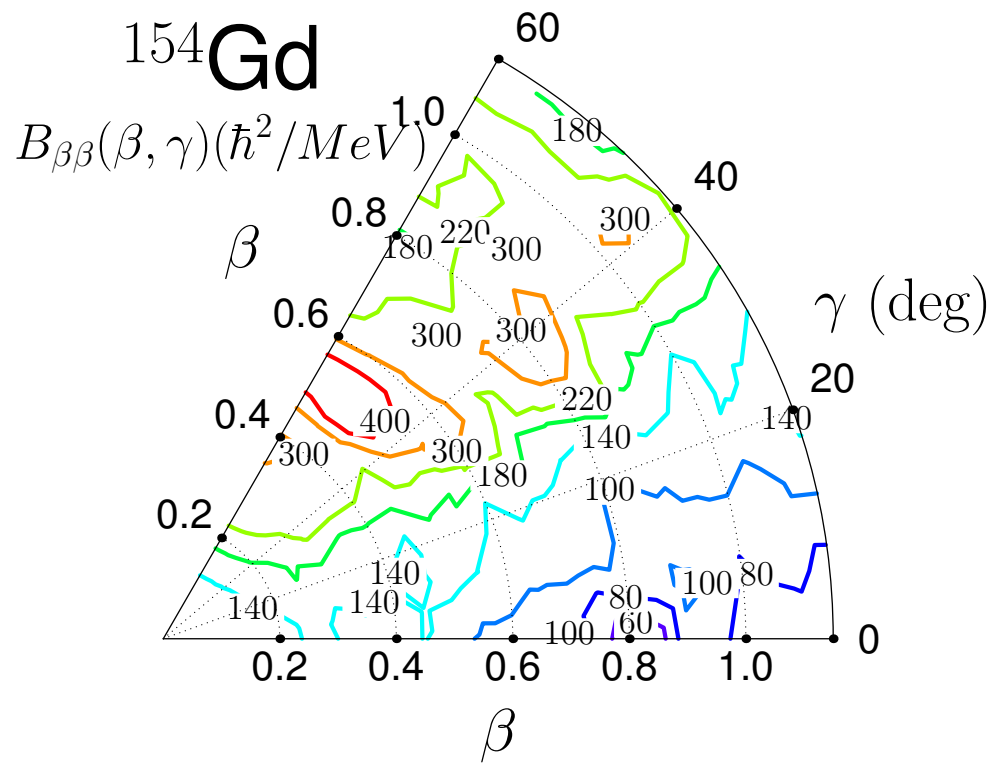}
\caption{(Color online) The cranking mass parameter $B_{\beta\beta}$
of $^{154}$Gd in the
$\beta - \gamma$ plane ($0\le \gamma\le 60^0$).}
\label{Fig8}
\end{figure}

\begin{figure}
\includegraphics[scale=1.1]{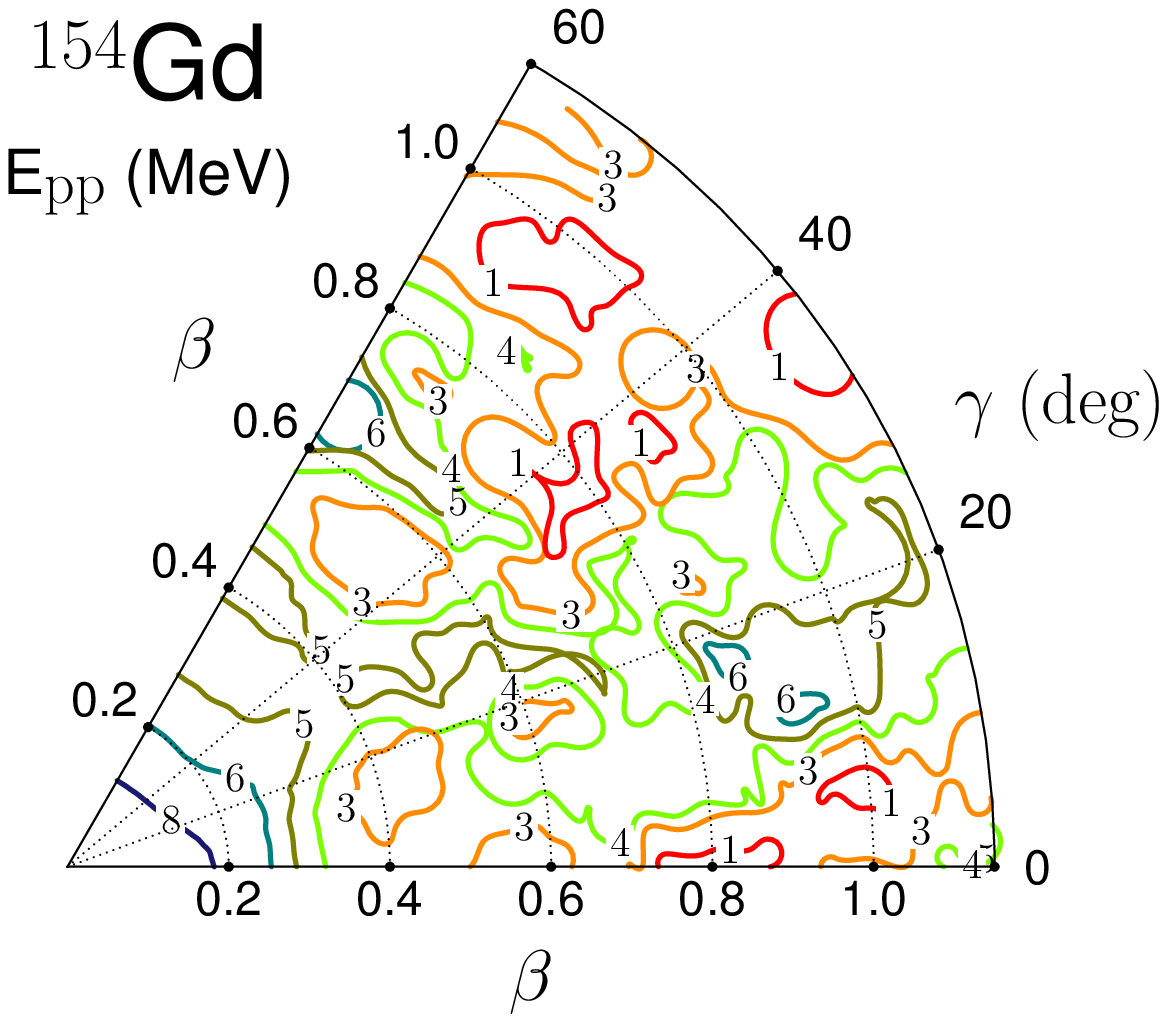}
\caption{(Color online) Proton pairing energy
of $^{154}$Gd in the
$\beta - \gamma$ plane ($0\le \gamma\le 60^0$). 
The contours join points
on the surface with the same energy (in MeV).}
\label{Fig9}
\end{figure}

\begin{figure}
\includegraphics[scale=1.1]{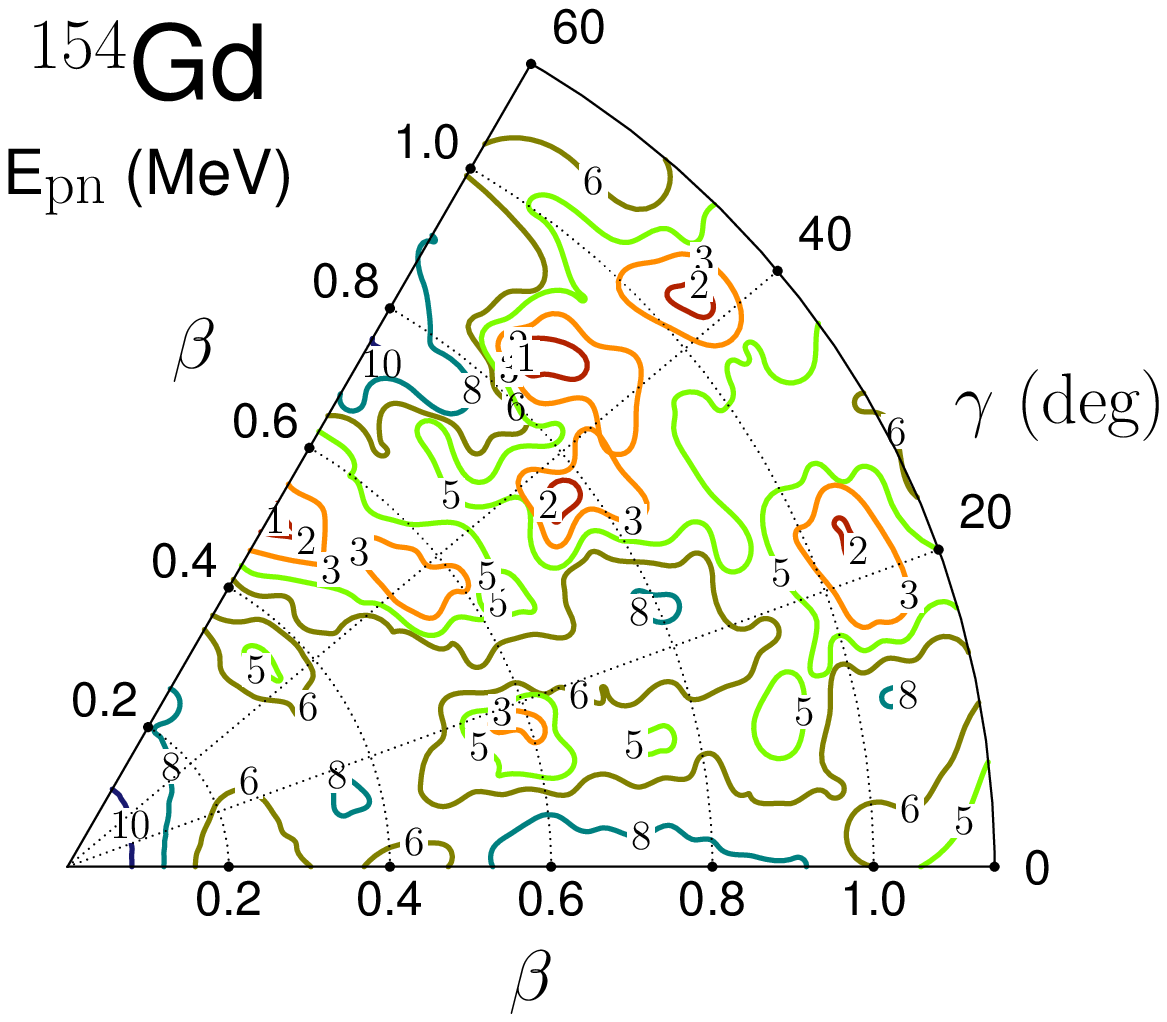}
\caption{(Color online) Neutron pairing energy
of $^{154}$Gd in the
$\beta - \gamma$ plane ($0\le \gamma\le 60^0$). 
The contours join points
on the surface with the same energy (in MeV). }
\label{Fig10}
\end{figure}

Starting from constrained self-consistent solutions, i.e. using the
single-particle wave functions, occupation probabilities, and
quasiparticle energies that correspond to each point on the energy
surfaces shown in Figs.~\ref{Fig2}-\ref{Fig6}, the parameters that
determine the collective Hamiltonian: mass parameters
$B_{\beta\beta}$, $B_{\beta\gamma}$, $B_{\gamma\gamma}$, three
moments of inertia $\mathcal{I}_k$, as well as the zero-point energy
corrections, are calculated as functions of the deformations $\beta$
and  $\gamma$. As an illustration, for $^{154}$Gd  the contour map 
of the inertia parameter $B_x$ (cf. Eq.~(\ref{Bx})) is shown
in Fig.~\ref{Fig7}, and the mass parameter $B_{\beta\beta}$ 
in Fig.~\ref{Fig8}. The inertia parameter decreases with the 
increase of the axial deformation $\beta$, but the  
dependence on $\gamma$ is very weak. The mass parameters,
on the other hand, display a pronounced dependence 
on both intrinsic deformations $\beta$ and  $\gamma$. 
We notice that $B_{\beta\beta}$ generally increases with 
the increase of $\gamma$ from prolate ($\gamma\approx 0^0$) toward 
oblate ($\gamma\approx  60^0$) configurations. The somewhat erratic 
behavior of $B_{\beta\beta}$, in particular, is mainly caused by the 
fluctuations of pairing correlations as function of $\beta$ and  $\gamma$ 
(cf. Eq.~(\ref{masspar-M}) with its strong dependence on the 
quasiparticle energies). This effect is illustrated in Figs.~\ref{Fig9} and 
\ref{Fig10}, where we plot the contour maps of the proton and neutron 
pairing energies, respectively, in the $\beta - \gamma$ plane. 
The fluctuations of pairing energies reflect the underlying shell 
structure and, because pairing correlations are described in the 
BCS approximations, pairing is strongly reduced wherever the 
level density around the Fermi level is small. As a result, mass 
parameters are locally enhanced in regions of weak pairing. 

\begin{figure}
\includegraphics[scale=1.0]{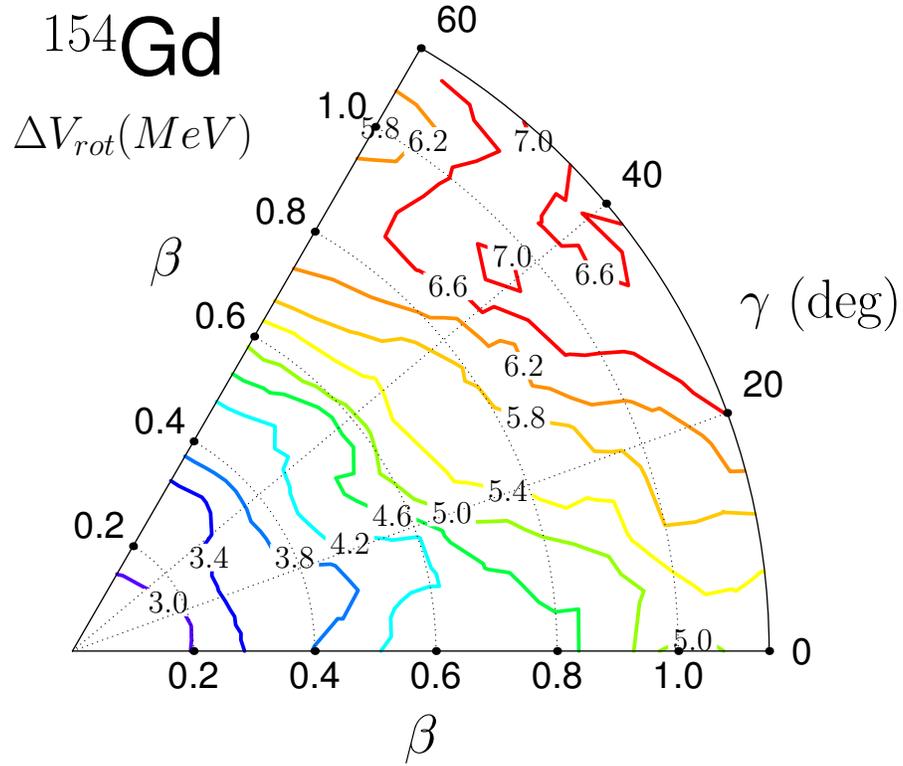}
\caption{(Color online) The rotational zero-point energy
of $^{154}$Gd in the
$\beta - \gamma$ plane ($0\le \gamma\le 60^0$).}
\label{Fig11}
\end{figure}

\begin{figure}
\includegraphics[scale=1.0]{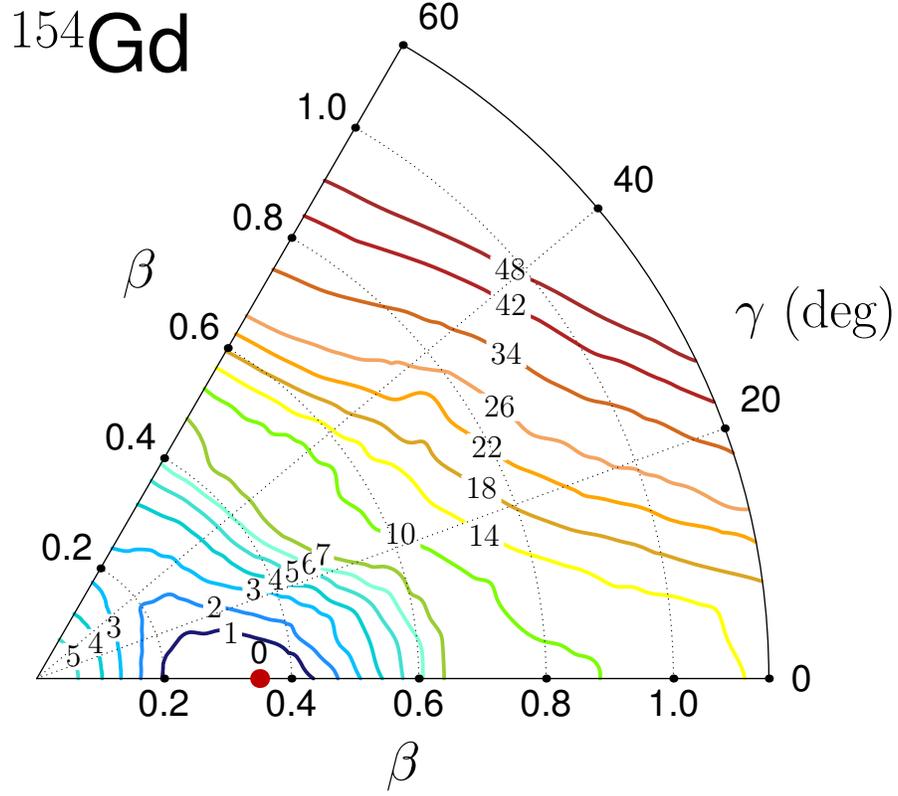}
\caption{(Color online) The potential $V_{\textnormal{coll}}$ 
(Eq.~(\ref{Vcoll})) of $^{154}$Gd in the
$\beta - \gamma$ plane ($0\le \gamma\le 60^0$). 
The contours join points
on the surface with the same energy (in MeV).}
\label{Fig12}
\end{figure}

Fig.~\ref{Fig11} displays the contour plot of the rotational zero 
point energy correction Eq.~(\ref{ZPE-rot}) for $^{154}$Gd.  
The rotational ZPE, of course, increases with the axial deformation 
$\beta$, but we also notice a pronounced dependence on $\gamma$.
The ZPE corrections are of the order of several MeV even in the region close
to the minimum and can, therefore, present a significant contribution to
the potential of the collective Hamiltonian. For $^{154}$Gd this is illustrated 
in Fig.~\ref{Fig12}, where we plot the potential $V_{\textnormal{coll}}$ 
(Eq.~(\ref{Vcoll})) in the $\beta - \gamma$ plane. When this plot is 
compared to the total mean-field energy Eq.~(\ref{etot}) (cf. Fig.~\ref{Fig3}), 
one notes that the main effect of ZPE corrections is to shift the 
position of the minimum to a larger prolate deformation, and to modify the 
shape of the potential around the minimum.

\begin{figure}
\includegraphics[scale=0.7,angle=270]{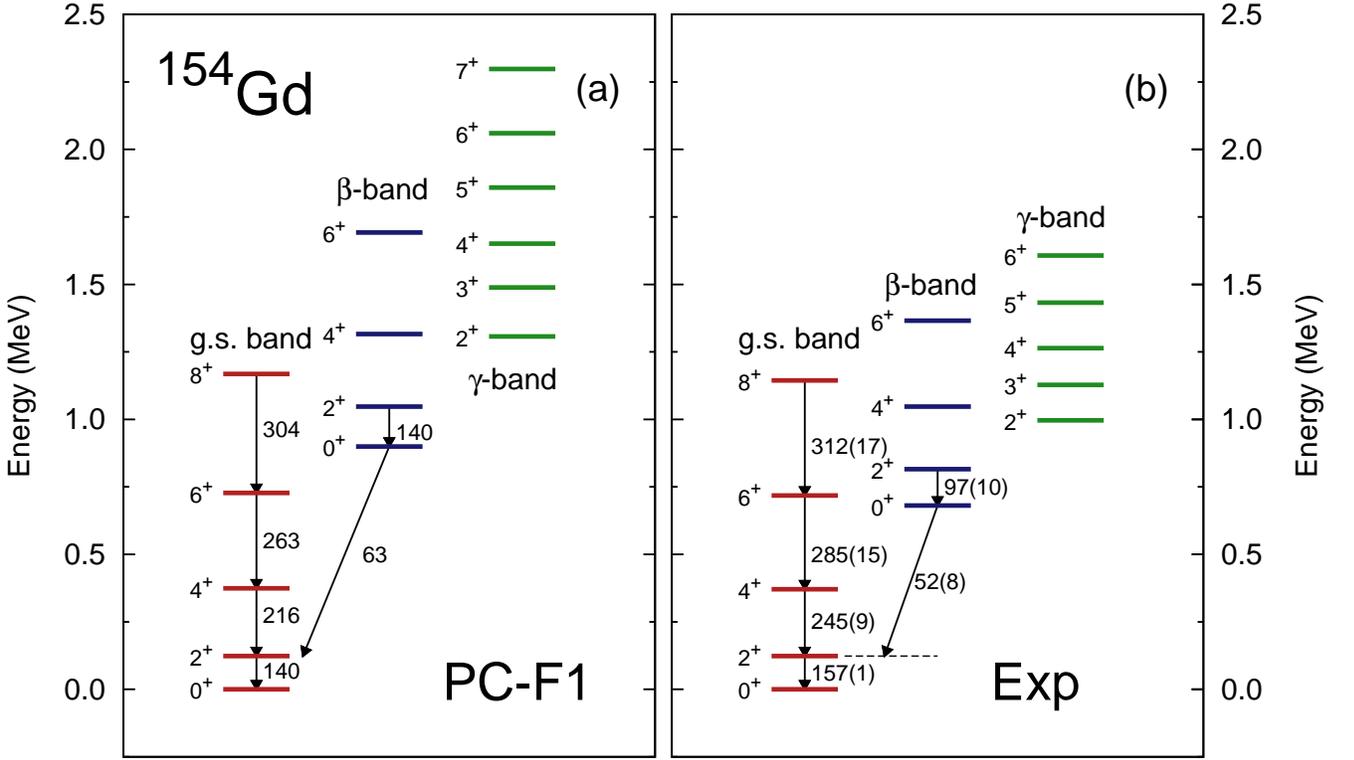}
\caption{(Color online) The level scheme of $^{154}$Gd calculated with the
PC-F1 relativistic density functional, in comparison with the experimental 
data~\cite{Gd154.exp}. 
The relative excitation energies are scaled by the common factor
$\approx 0.69$, adjusted to the experimental energy of the state $2_1^+$. 
The $B(E2)$ values are given in Weisskopf units.}
\label{Fig13}
\end{figure}

The diagonalization of the resulting Hamiltonian yields the
excitation energies and the collective wave functions for each value
of the total angular momentum and parity $I^\pi$. In addition to the
yrast ground-state band, in deformed and transitional nuclei excited
states are usually also assigned to (quasi) $\beta$ and $\gamma$
bands. This is done according to the distribution of the angular
momentum projection $K$ quantum number defined in Eq. (\ref{NK}).
Excited states with predominant $K=2$ components in the wave function
are assigned to the $\gamma$-band, whereas the $\beta$-band comprises
states above the yrast characterized by dominant $K=0$ components. As
an example, in Fig.~\ref{Fig13} we display the PC-F1 excitation
spectrum of $^{154}$Gd, in comparison with available data
~\cite{Gd154.exp}. The relative excitation energies within all three
theoretical bands are scaled by the common factor $\approx 0.69$,
determined in such a way that the calculated energy of the $2_1^+$
state coincides with the experimental value. This leaves the
bandheads of the $\gamma$ and $\beta$ bands unaltered, and
corresponds to an enhancement of the effective moment of inertia by
$\approx 45\%$. The scaling of the relative excitation energies is
introduced because of the well known fact that the Inglis-Belyaev
(IB) formula (\ref{Inglis-Belyaev}) predicts effective moments of
inertia that are considerably smaller than empirical values. More
realistic values are only obtained if one uses the Thouless-Valatin
(TV) formula~\cite{TV.62}, but this procedure is computationally much
more demanding, and it has not been implemented in the current
version of the model. Here we rather follow the prescription of
Ref.~\cite{LGD.99} where, by comparing the TV and IB  moments of
inertia as functions of the axial deformation for superdeformed bands
in the $A=190-198$ mass region, it was shown that the
Thouless-Valatin correction to the perturbative expression IB is
almost independent of deformation, and does not include significant
new structures in the moments of inertia. It was thus suggested that
the moments of inertia to be used in the collective Hamiltonian can
be simply related to the IB values through the minimal prescription:
$\mathcal{I}_k (q) = \mathcal{I}^{IB}_k (q) (1+ \alpha)$, where $q$
denotes the generic deformation parameter, and $\alpha$ is a constant
that can be determined in a comparison with data. The value of
$\alpha \approx 0.45$ used for the excitation spectrum of $^{154}$Gd
is somewhat larger than the values determined in the mass $A=190-198$
region~\cite{LGD.99}.

\begin{figure}
\includegraphics[scale=0.7]{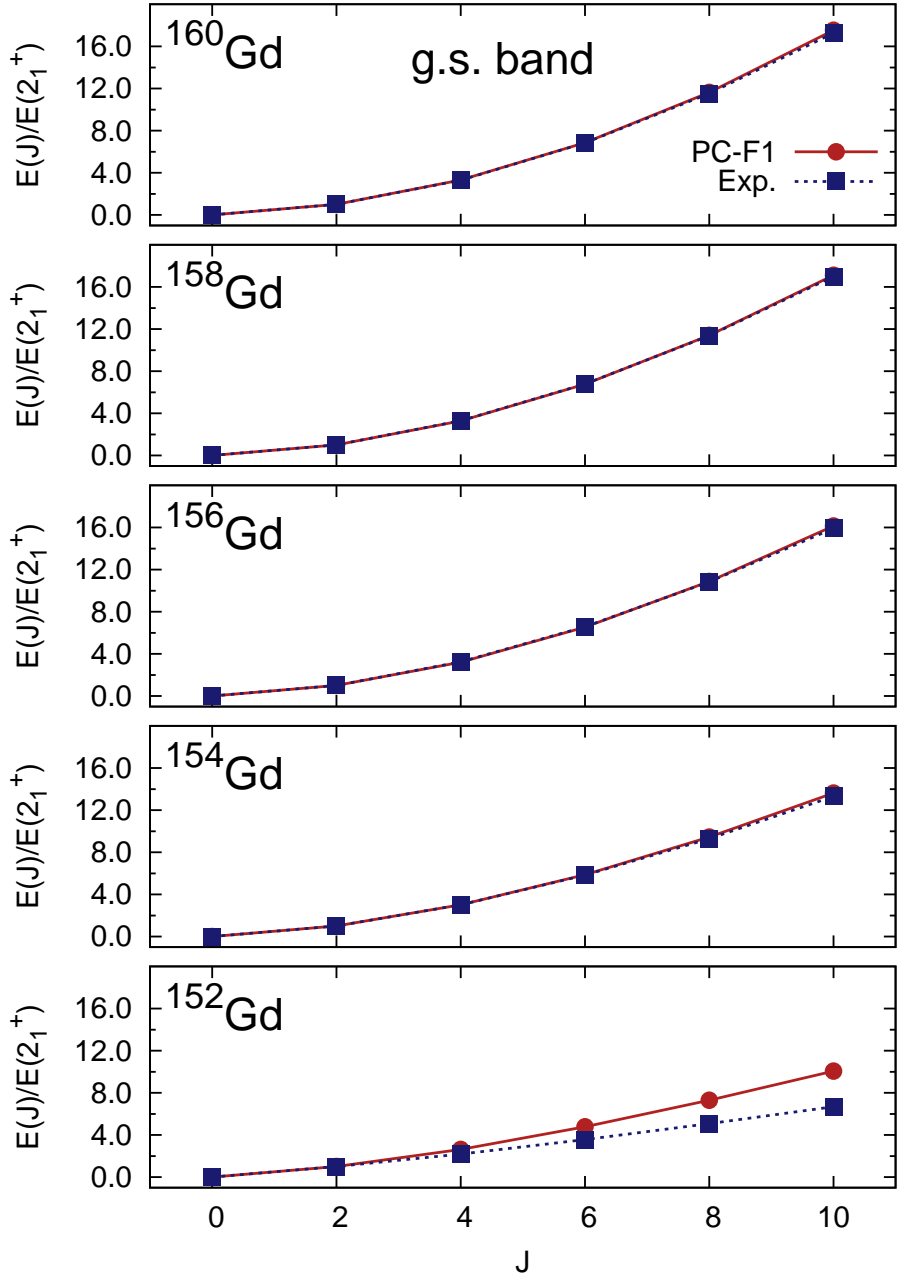}
\caption{(Color online) Relative ground-state band excitation energies 
in $^{152-160}$Gd. For each nucleus the theoretical energies 
are scaled by a common factor, adjusted to the experimental 
energy of the $2_1^+$ state.}
\label{Fig14}
\end{figure}

\begin{figure}
\includegraphics[scale=0.7]{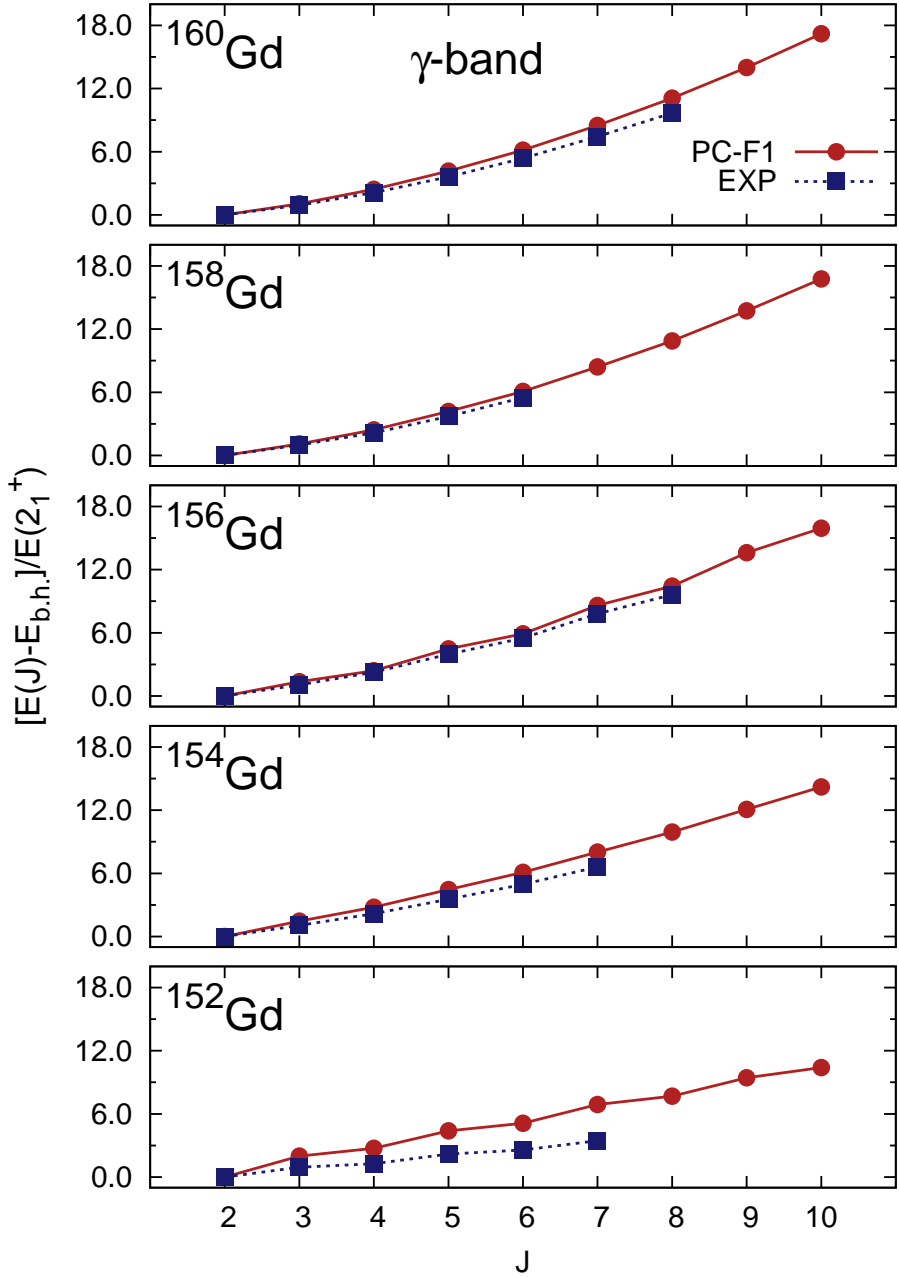}
\vspace{1cm}
\caption{(Color online) Comparison between the theoretical and
experimental $\gamma$-band excitation energies for $^{152-160}$Gd. 
The scaling factors are the same as for the ground-state bands in 
Fig.~\ref{Fig14}.}
\label{Fig15}
\end{figure}

\begin{figure}
\includegraphics[scale=0.7]{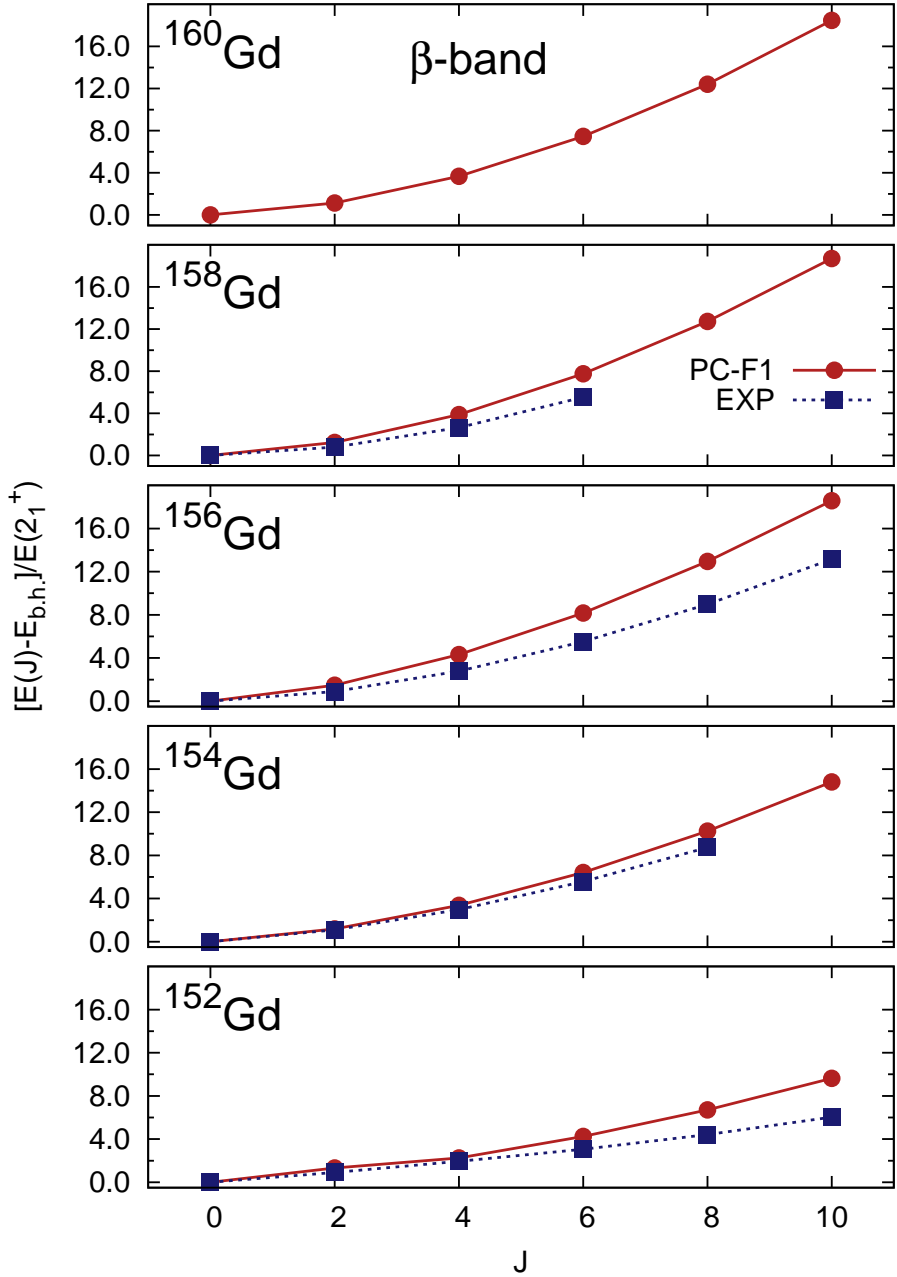}
\vspace{1cm}
\caption{(Color online) Same as in Fig.~\ref{Fig15}, but for the
$\beta$-bands in $^{152-160}$Gd.
The scaling factors are the same as for the ground-state bands in 
Fig.~\ref{Fig14}.}
\label{Fig16}
\end{figure}

When the IB effective moments of inertia are renormalized to the
empirical values by scaling the relative excitation energies to
reproduce the experimental position of the state $2_1^+$, the
resulting bands are in good agreement with data. This is
illustrated in Figs.~\ref{Fig14}-\ref{Fig16},  where we plot the
excitation energies with respect to bandheads, for the ground-state
band, $\gamma$, and $\beta$ bands in $^{152-160}$Gd, respectively. 
For each nucleus the relative excitation energies within the three bands 
are scaled by a common factor, adjusted to the experimental 
energy of the $2_1^+$ state, as explained above. These factors are 
rather similar for four isotopes: 0.69 for $^{154}$Gd, 0.67 for $^{156}$Gd, 
0.69 for $^{158}$Gd, and 0.72 for $^{160}$Gd. An exception is 
the lightest Gd isotope considered here: $^{152}$Gd, for which 
this factor is actually 1.08, i.e. the theoretical $2_1^+$ state 
($E_x = 0.318$ MeV) is slightly below the experimental $2_1^+$ state 
($E_x = 0.344$ MeV). However, as shown in Figs.~\ref{Fig14}-\ref{Fig16}, 
for this nucleus the calculated ground-state band, as well as the 
(quasi) $\gamma$ and $\beta$ bands, do not follow 
very closely the experimental spectra. The deviation from the
experimental trend at higher angular momenta  can
probably be explained by the fact that this is a weakly-deformed
transitional nucleus for which the assumption of a constant moment of
inertia, implicit in the expression $\mathcal{I}_k (q) =
\mathcal{I}^{IB}_k (q) (1+ \alpha)$, and of relatively pure $\beta$
and $\gamma$ bands, does not present a very good approximation.

\begin{figure}
\includegraphics[scale=0.7]{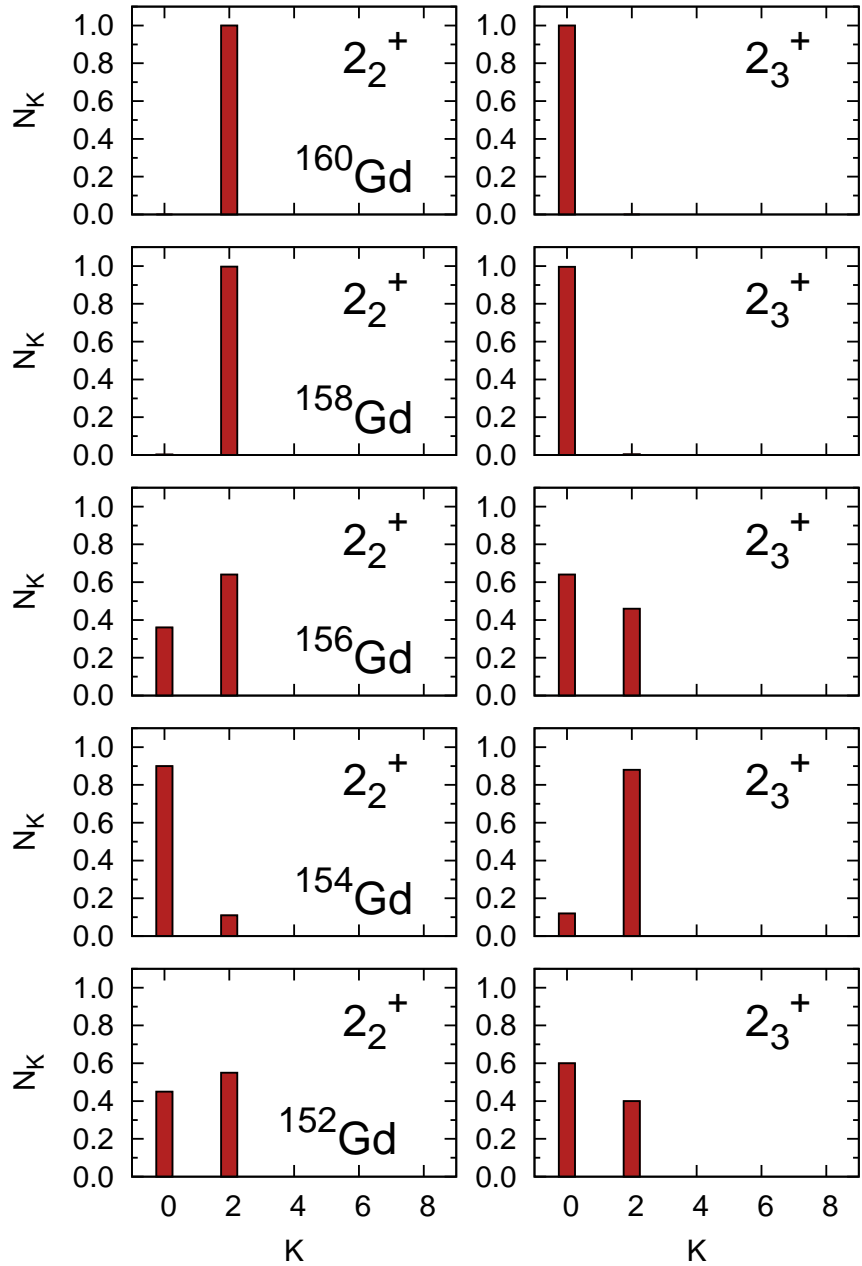}
\caption{(Color online) Distribution of the $K$-components (projection of the
angular momentum on the body-fixed symmetry axis) in the
collective wave functions for the states: $2_2^+$ and $2_3^+$.}
\label{Fig17}
\end{figure}

\begin{figure}
\includegraphics[scale=0.7]{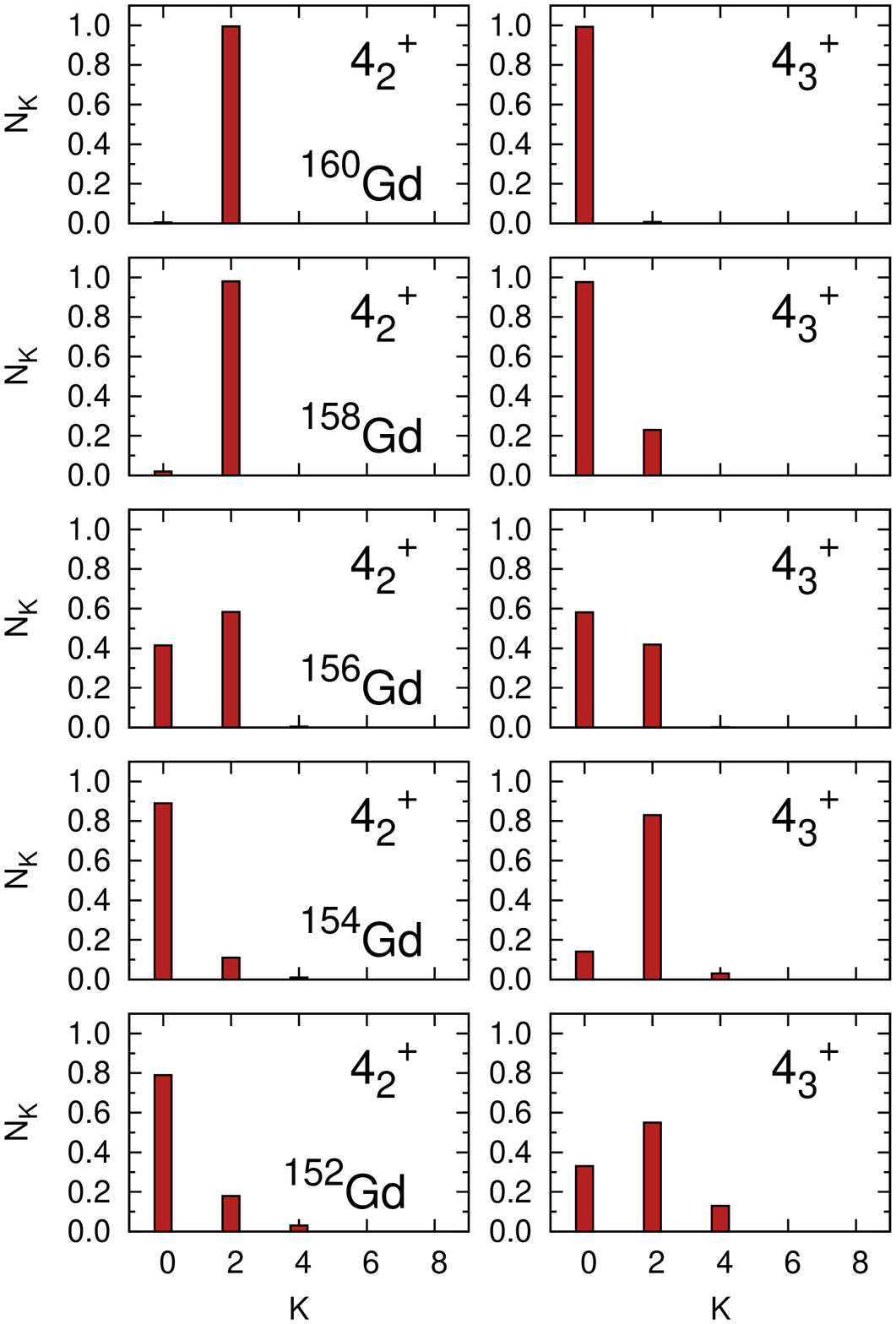}
\caption{(Color online) Same as in Fig.~\ref{Fig17}, but for the
states: $4_2^+$ and $4_3^+$.}
\label{Fig18}
\end{figure}

\begin{figure}
\includegraphics[scale=0.7]{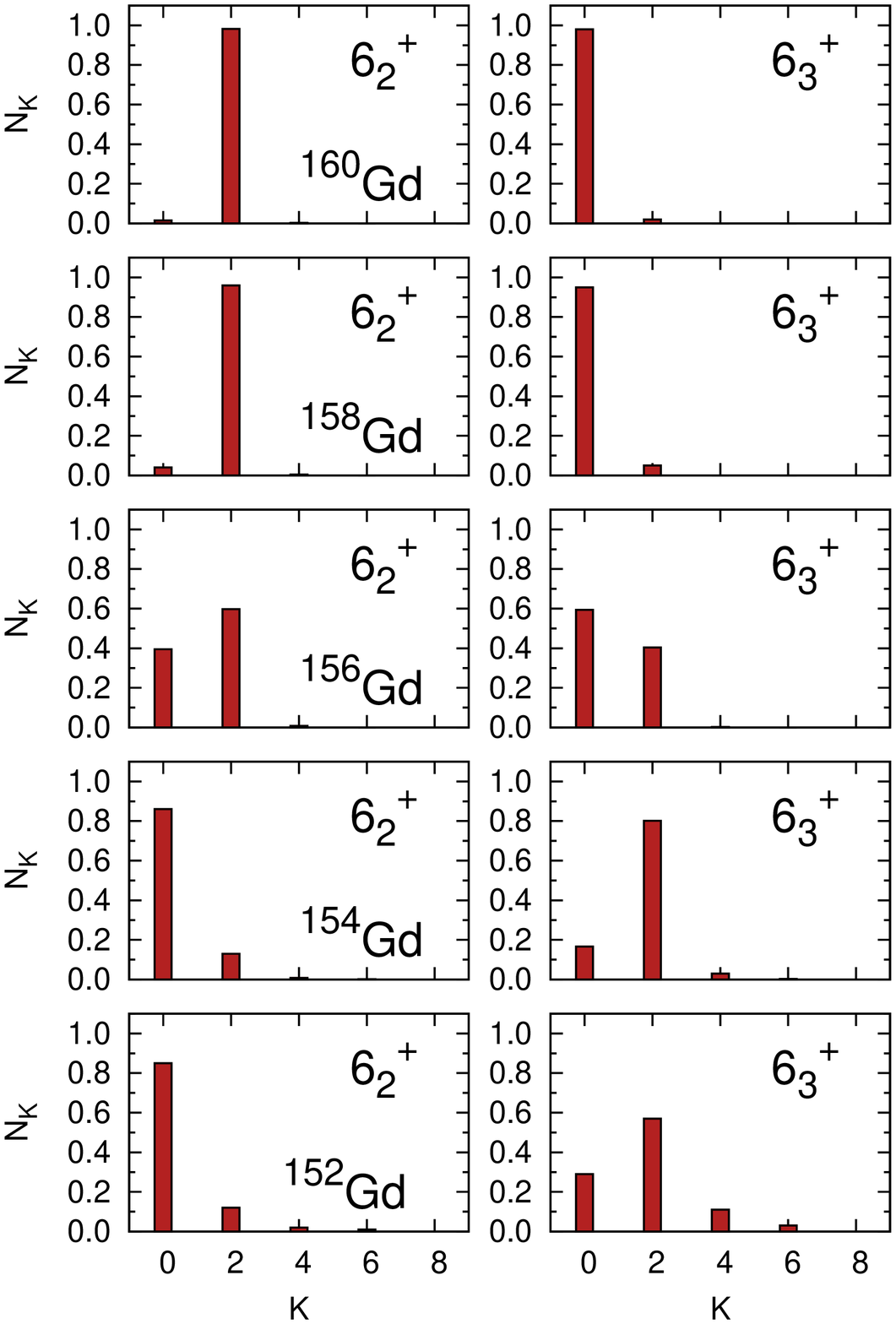}
\caption{(Color online) Same as in Fig.~\ref{Fig17}, but for the
states: $6_2^+$ and $6_3^+$.}
\label{Fig19}
\end{figure}

\begin{figure}
\includegraphics[scale=0.7]{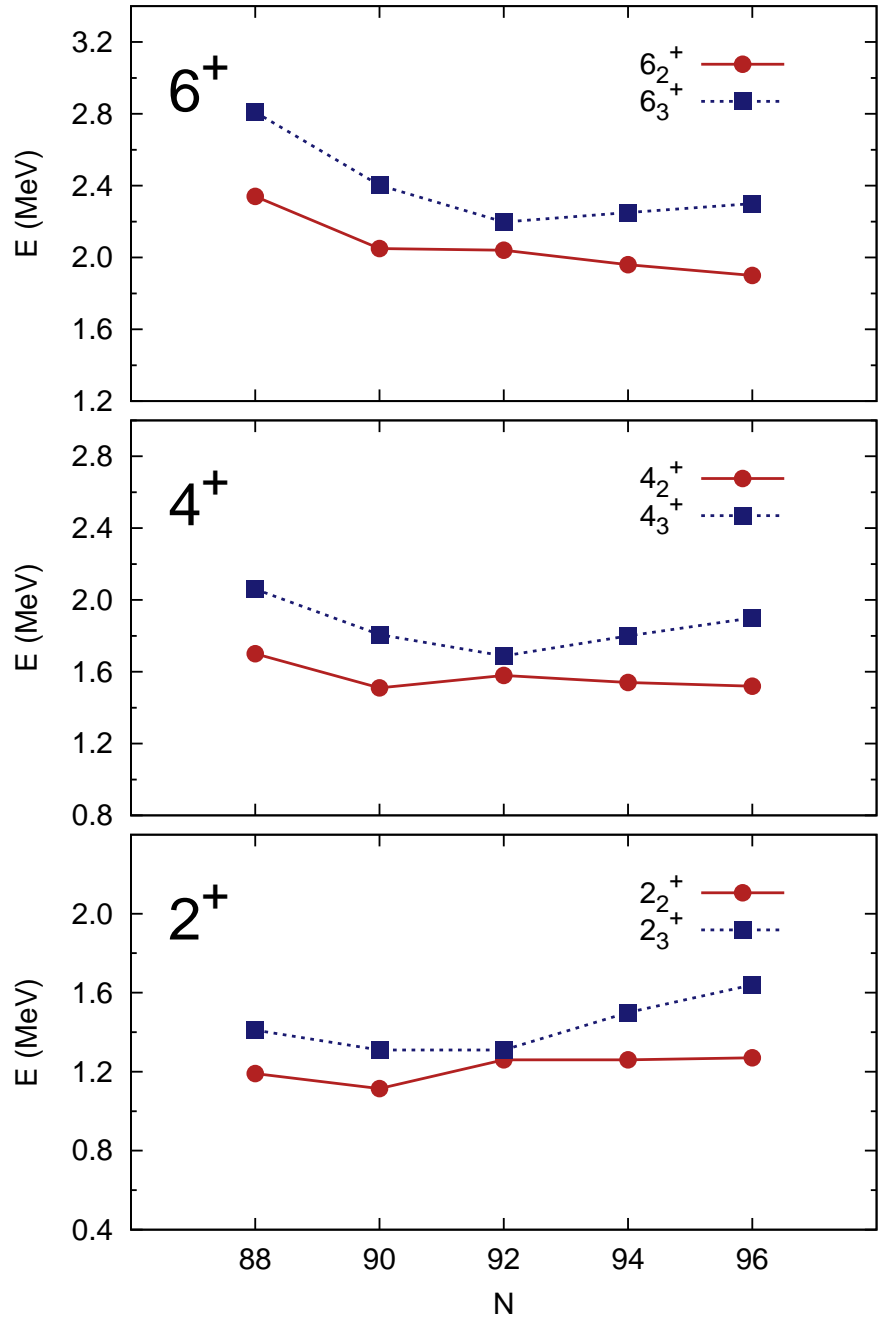}
\caption{(Color online) Excitation energies of the second and third
states: $2^+$, $4^+$ and $6^+$ in the Gd isotopic chain,
as functions of the neutron number.}
\label{Fig20}
\end{figure}

The calculated $\beta$ bands are compared with available data in
Fig.~\ref{Fig16}. In comparison with the $\gamma$ bands (cf.
Fig.~\ref{Fig15}), the agreement with data is better in $^{152}$Gd,
but the deviation from experiment is more pronounced in $^{156}$Gd.
To understand in more detail the discrepancy between the calculated
and empirical $\beta$ and $\gamma$ bands, we need to consider the
distribution of the angular momentum projection $K$ quantum number
(cf. Eq.~(\ref{NK})) in these bands. As explained above, the
calculated second and third eigenstate for each angular momentum are
assigned either to the $\beta$ or $\gamma$ band, on the basis of the
predominant $K=0$ or $K=2$ components, respectively. The
distributions of $K$ components in the wave functions of the
calculated second and third $2^+$, $4^+$ and $6^+$ states are plotted
in Figs.~\ref{Fig17}-\ref{Fig19}, respectively. In the case of
$^{152}$Gd, in particular, we find a pronounced mixing of the $K=0$
or $K=2$ components and, with increasing angular momentum,
contributions of higher-$K$ components are present in the wave
functions. This is consistent with the observation that $^{152}$Gd is
a transitional nucleus and, therefore, excited states can only
approximately be assigned to (quasi) $\beta$ and $\gamma$ bands.
Increasing the neutron number toward heavier and more deformed Gd
isotopes, the distributions of $K$ components become sharp and states
can unambiguously be grouped into $\beta$ and $\gamma$ bands. This
is, of course, characteristic for well deformed nuclei. One
exception, however, is the calculated spectrum of $^{156}$Gd, where
we find significant mixing of $K=0$ and $K=2$ components in the wave
functions of second and third $2^+$, $4^+$ and $6^+$ states, as well
as for higher angular momenta. The more pronounced mixing between the
$\beta$ and $\gamma$ bands occurs because,  in this particular
isotope, the calculated second and third even-spin states are almost
degenerate in energy,  as shown in Fig.~\ref{Fig20}.

\begin{figure}
\includegraphics[scale=0.7]{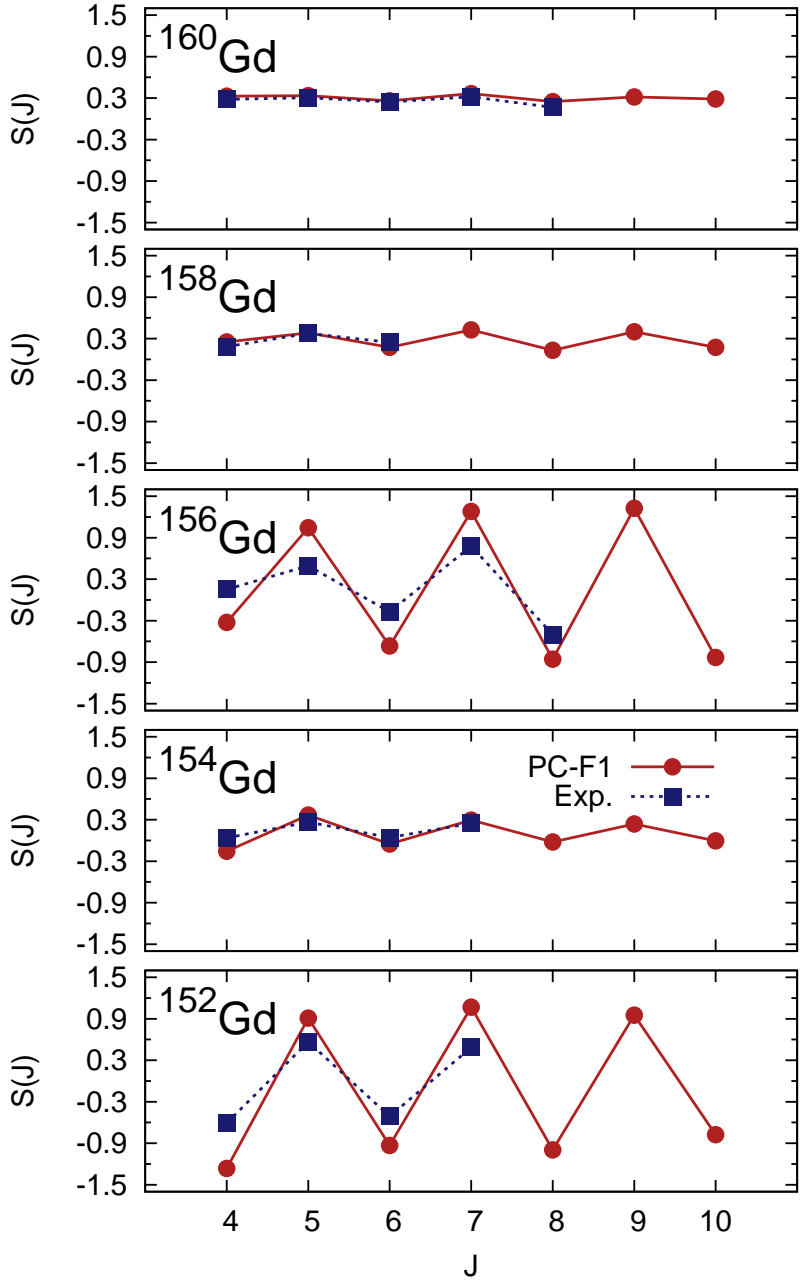}
\vspace{1cm}
\caption{(Color online) Staggering $S(J)$ [Eq.~(\ref{staggering})] in
the $\gamma$-bands of $^{152-160}$Gd. Theoretical predictions are
compared with experimental values.}
\label{Fig21}
\end{figure}

The level of $K$-mixing is reflected in the staggering in energy
between odd- and even-spin states in the (quasi) $\gamma$-bands (cf.
Fig.~\ref{Fig15}). The staggering can be quantified by considering the
differential quantity~\cite{ZC.91}
\begin{equation}
S(J)=\frac{\left\{E[J_\gamma^+] - E[(J-1)_\gamma^+]\right\} 
       - \left\{E[(J-1)_\gamma^+] - E[(J-2)_\gamma^+]\right\}}{E[2_1^+]},
\label{staggering}
\end{equation}
which measures the displacement of the $(J-1)_\gamma^+$ level
relative to the average of its neighbors, $J_\gamma^+$ and
$(J-2)_\gamma^+$, normalized to the energy of the first excited state
of the ground state band, $2_1^+$. Because of its differential form,
$S(J)$ is very sensitive to structural changes. For an axially
symmetric rotor $S(J)$ is, of course, constant. In a nucleus with a
deformed $\gamma$-soft potential, $S(J)$ oscillates between negative
values for even-spin states and positive values for odd-spin states,
with the magnitude slowly increasing with spin. For a triaxial
potential the level clustering in the (quasi) $\gamma$-band is
opposite, and $S(J)$ oscillates between positive values for even-spin
states and negative values for odd-spin states. In this case the
magnitude of $S(J)$ increases more rapidly with spin, as compared to
the  $\gamma$-soft potential. In a recent study of staggering of
$\gamma$-band energies and the transition between different
structural symmetries in nuclei~\cite{CBZC.07}, the experimental
energy staggering in $\gamma$-bands of several isotopic chains has
been investigated as a signature for the $\gamma$ dependence of the
potential. In Fig.~\ref{Fig21} we plot the staggering in the
$\gamma$-band  for the chain of the Gd isotopes, calculated with the
PC-F1 relativistic density functional. One notices how the pronounced
$K$-mixing in $^{152}$Gd and $^{156}$Gd (cf. Figs.
~\ref{Fig17}-\ref{Fig19}) leads to the strong staggering observed in
the corresponding (quasi) $\gamma$-bands. The calculation reproduces
both the empirical oscillatory behavior and, with the exception of
low-spin states in $^{156}$Gd, also the magnitude of $S(J)$. Starting
from the $\gamma$-soft $^{152}$Gd (negative values for even-spin
states and positive values for odd-spin states), $S(J)$ evolves
toward the axially symmetric rotor limit ($S(J)=0.33$) in $^{158}$Gd
and $^{160}$Gd.

The assignment of even-spin states above the yrast either to the $\beta$ or 
$\gamma$ bands, on the basis of the predominant $K=0$ or $K=2$ components
(Figs.~\ref{Fig17}-\ref{Fig19}), and the level of K-mixing inferred from the 
differential quantity $S(J)$ Eq.~(\ref{staggering}) (Fig.~\ref{Fig21}), has a 
correspondence in the calculated average values of 
the deformation parameters $\beta$ and 
$\gamma$ (cf. Eqs.~(\ref{avbeta}) and (\ref{avgamma})). In Table~\ref{Tab-B} 
we collect the average $\beta$ and $\gamma$ deformations for the 
calculated first, second, and third $2^+$, $4^+$ and $6^+$ states in 
$^{152-160}$Gd. For those nuclei where the K-mixing is weak (sharp 
distribution of K-components in Figs.~\ref{Fig17}-\ref{Fig19}, weak staggering 
of $S(J)$ in Fig.~\ref{Fig21}): $^{154}$Gd, $^{158}$Gd, and $^{160}$Gd, 
the average values of the deformation parameters are almost identical 
for states belonging to the ground-state band and those assigned to the 
$\beta$-band. States assigned to the $\gamma$-band ($K =2$) are 
consistently characterized by larger average values of the angle $\gamma$. 
This distinction does not appear in the spectra of the two nuclei for which the 
model predicts pronounced K-mixing: $^{152}$Gd and $^{156}$Gd. 

\begin{figure}
\includegraphics[scale=0.7]{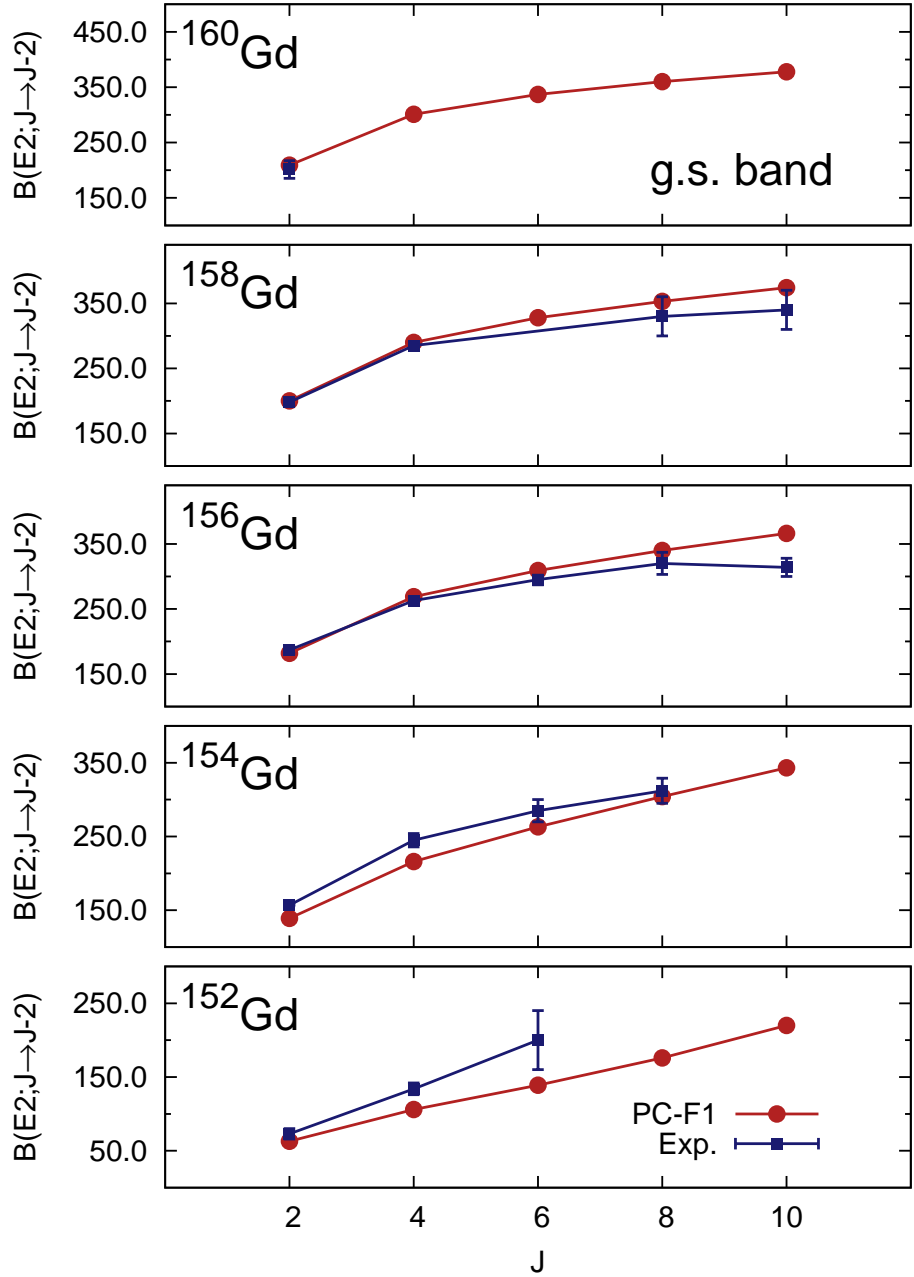}
\vspace{1cm}
\caption{(Color online) $B(E2)$ values (in Weisskopf units) for the
ground-state band transitions $J_1^+\to (J-2)_1^+$ in $^{152-160}$Gd.
Theoretical values calculated with the PC-F1 relativistic density
functional are compared with data.}
\label{Fig22}
\end{figure}

An important advantage of using structure models based on
self-consistent mean-field single-particle solutions is the fact that
physical observables, such as transition probabilities and
spectroscopic quadrupole moments, are calculated in the full
configuration space and there is no need for effective charges. Using
the bare value of the proton charge in the electric quadrupole
operator $\mathcal{\hat{M}}(E2)$, the transition probabilities between
eigenvectors of the collective Hamiltonian can be directly compared
with data. In addition to the calculated energy spectrum of
$^{154}$Gd, in Fig.~\ref{Fig13} we have also compared the resulting
B(E2) values (in Weisskopf units) for transitions within the
ground-state band, and the transitions $2_\beta^+ \rightarrow
0_\beta^+$ and $0_\beta^+ \rightarrow 2_{\rm g.s.}^+$, with available
experimental values. The agreement between theoretical B(E2) values
and data is very good, especially considering that the calculation of
transition probabilities is completely parameter-free. We also notice
the remarkable prediction for the interband transition $0_\beta^+
\rightarrow 2_{\rm g.s.}^+$, in excellent agreement with experiment.
Finally, in Fig.~\ref{Fig22} we plot the calculated $B(E2)$ values (in
Weisskopf units) for the ground-state band transitions $J_1^+\to
(J-2)_1^+$ in $^{152-160}$Gd, together with the available
experimental values. The model clearly reproduces the empirical trend
of ground-state band transitions in Gd isotopes and, except perhaps
for the transitional nucleus $^{152}$Gd, the theoretical predictions
are in excellent agreement with data even for higher angular momentum
states.

\section{\label{secIV}Summary and Outlook}

To describe complex excitation patterns and electromagnetic
transition rates associated with the evolution of shell structures
starting from stable nuclei and extending toward regions of exotic
short-lived systems far from $\beta$-stability, nuclear structure
methods must be developed that are based on a universal microscopic
framework. Properties of a vast majority of nuclides with a large number of
valence nucleons are best described by nuclear energy
density functionals. However, for a quantitative description of energy spectra and 
transition probabilities one must be able to go beyond the lowest 
order in which the EDFs are implemented -- the mean-field approximation, 
and systematically include correlations related to the restoration of broken 
symmetries and to fluctuations of collective variables. In the framework 
of non-relativistic EDFs, in recent years several models have been developed 
that use the generator coordinate method (GCM) to perform configuration 
mixing calculations with angular-momentum and particle-number projected 
mean-field (SR EDF) states. In most applications the calculations have 
been restricted to axially symmetric, parity conserving configurations. 

In a recent series of papers, of which the present is the third part,
we have extended the relativistic EDFs to include symmetry
restoration and fluctuations of quadrupole deformations. While in the
first two parts the GCM was used in configuration mixing calculations
with axially symmetric relativistic wave functions, this work has been focused on the
description of general triaxial shapes. We have developed a new implementation
for the solution of the eigenvalue problem of a five-dimensional
collective Hamiltonian for quadrupole vibrational and rotational
degrees of freedom, with parameters determined by constrained
self-consistent relativistic mean-field calculations for triaxial
shapes. In addition to the
self-consistent mean-field potential of the PC-F1 relativistic
density functional in the particle-hole channel, for open-shell
nuclei pairing correlations are included in the BCS approximation.
The resulting single-nucleon wave functions, energies and occupation
factors, as functions of the quadrupole deformations, provide the
microscopic input for the parameters of the collective Hamiltonian:
three mass parameters: $B_{\beta\beta}$, $B_{\beta\gamma}$,
$B_{\gamma\gamma}$, three moments of inertia $\mathcal{I}_k$,  and
the collective potential including zero-point vibrational and
rotational energy corrections. The moments of inertia are calculated
using the Inglis-Belyaev formula, and the mass parameters associated
with the quadrupole collective coordinates are determined in the
cranking approximation. 
An extensive test has been carried out in calculations of
potential energy surfaces, and the resulting collective excitation
spectra and transition probabilities, for the chain of even-even
gadolinium isotopes. Results for excitation energies in ground-state,
(quasi) $\beta$ and $\gamma$ bands, and the corresponding interband
and intraband transition probabilities have been compared with
available data on even-even Gd isotopes: $^{152-160}$Gd. 

There are some obvious improvements that need to be implemented 
in the model. For instance, because the
Inglis-Belyaev formula gives effective moments of inertia that are
lower than empirical values, all the calculated relative
excitation energies in $^{152-160}$Gd had to be scaled with respect
to the experimental energy of the $2_1^+$ states. The moments of
inertia can be improved by
including the Thouless-Valatin dynamical rearrangement contributions.
For the rotational degrees of freedom for which the collective
momenta are known, the inertia parameters can be obtained from the 
solutions of cranked RMF equations. For the deformation
coordinates $q_0$ and $q_2$ the situation is more complicated,
because the corresponding momentum operators  $\hat{P}_0$ and
$\hat{P}_2$ have to be calculated from the solution of Thouless-Valatin 
equations \cite{TV.62} at each deformation point. 
Because cranking breaks time-reversal symmetry, in both cases the
inclusion of pairing correlations necessitates the extension of the 
model from the simple RMF+BCS to the full 
relativistic Hartree-Bogoliubov framework~\cite{VALR.05}. 

\appendix
\section{\label{app-A}Three-dimensional solution of the Dirac equation}

To solve the Dirac equation (\ref{dirac}) for triaxially
deformed potentials, the single-nucleon spinors are expanded
in the basis of eigenfunctions of a three-dimensional harmonic oscillator 
(HO) in Cartesian coordinates \cite{KR.88,DD.97,YaoII,PMR.08}.
In one dimension:
\begin{equation}
\phi_{n_\mu}(x_\mu) = b_\mu^{-1/2}H_{n_\mu}(\xi_\mu)e^{-\xi_\mu^2/2}\;,
\quad (\mu\equiv x,y,z)
\end{equation}
$\xi_\mu \equiv x_\mu/b_\mu$, and the oscillator length is defined as
\begin{equation}
b_\mu = \sqrt{\frac{\hbar}{m\omega_\mu}}\; .
\end{equation}
$H_n(\xi)$ denotes the normalized Hermite polynomials
\begin{equation}
\label{orthogonality}
\int_{-\infty}^{\infty}{H_n(\xi)H_{n^\prime}(\xi)e^{-\xi^2}d\xi} = \delta_{nn^\prime}\;.
\end{equation}
The basis state can be defined as the product of three HO wave functions
(one for each dimension) and the spinor:
\begin{equation}
\Phi_{\alpha}(\mathbf{r};m_s) =
    \phi_{n_x}(\xi_x) \phi_{n_y}(\xi_y)\phi_{n_z}(\xi_z)\chi_{m_s},
\label{product}
\end{equation}
where the notation is: $\alpha \equiv \{n_x,n_y,n_z\}$. It will be
more convenient to use the eigenstates of the x-simplex operator
defined by the relation
\begin{equation}
\hat{S}_x = \hat{P}e^{-i\pi J_x},
\end{equation}
where $\hat{P}$ denotes the parity operator.
It is easily verified that the x-simplex operator acting on the state
$\Phi_{\alpha}(\mathbf{r};m_s)$  leads to
\begin{equation}
\hat{S}_x\Phi_{\alpha}(\mathbf{r};m_s)  =
    -i(-1)^{n_x} \Phi_{\alpha}(x,y,z;-m_s)\;.
\end{equation}
The eigenstates of the x-simplex operator with positive and
negative eigenvalues read:
\begin{align}
\label{basis-pos}
\Phi_{\alpha}(\mathbf{r};+) &=  \phi_{n_x}(x) \phi_{n_y}(y)\phi_{n_{z}}(z)
     \frac{i^{n_y}}{\sqrt{2}}\left[ \chi_+ - (-1)^{n_x}\chi_-\right] \\
\label{basis-neg}
\Phi_{\alpha}(\mathbf{r};-) &=  \phi_{n_x}(x) \phi_{n_y}(y)\phi_{n_{z}}(z)
\frac{i^{n_y}}{\sqrt{2}} (-1)^{n_x+n_y+1}
\left[ \chi_+ + (-1)^{n_x}\chi_-\right] \;.
\end{align}

For the Dirac spinor with positive simplex eigenvalue, the large component
corresponds to positive, and the small component to negative eigenvalues
\begin{equation}
\psi_i(\mathbf{r},+) = \left( \begin{array}{c} f_i(\mathbf{r},+) \\
            ig_i(\mathbf{r},-) \end{array}  \right).
\end{equation}
The large and small component are expanded in terms of
the basis states Eqs.~(\ref{basis-pos}) and (\ref{basis-neg}):
\begin{equation}
f_i(\mathbf{r};+) =
    \sum_{\alpha}^{\alpha_{max}}{f_i^\alpha\Phi_{\alpha}(\mathbf{r};+)}
\quad \textnormal{and} \quad
g_i(\mathbf{r};-) =
    \sum_{\tilde{\alpha}}^{\tilde{\alpha}_{max}}
             {g_i^{\tilde{\alpha}}\Phi_{\tilde{\alpha}}(\mathbf{r};-)}\;.
\end{equation}
To avoid the occurrence of spurious states, $\alpha_{max}$ and
$\tilde{\alpha}_{max}$ are chosen in such a way that the
corresponding major quantum numbers $N=n_x+n_y+n_z$ are not larger
than some arbitrary $N_F$ for the expansion of large components, and
not larger than $N_F+1$ for the expansion of small components
~\cite{GRT.90}. The single-nucleon Dirac equation:
\begin{equation}
\left( \begin{array}{cc} V+m^* & \bm{\sigma} \cdot \bm{\nabla} \\
                                -\bm{\sigma} \cdot \bm{\nabla} & V-m^*
\end{array} \right)
\left( \begin{array}{c} f_i(\mathbf{r};+) \\ g_i(\mathbf{r};-)
       \end{array} \right)
= \epsilon_i  \left( \begin{array}{c} f_i(\mathbf{r};+) \\ g_i(\mathbf{r};-)
                     \end{array} \right),
\label{dirac-app}
\end{equation}
with the effective nucleon mass $m^* = m+S$,
reduces to a symmetric matrix eigenvalue problem
\begin{equation}
\left( \begin{array}{cc} \mathcal{A}_{\alpha\alpha^\prime} &
                         \mathcal{B}_{\alpha\tilde{\alpha}^\prime} \\
                         \mathcal{B}_{\tilde{\alpha}\alpha^\prime} &
                         \mathcal{C}_{\tilde{\alpha}\tilde{\alpha}^\prime}
\end{array} \right)
\left( \begin{array}{c} f_i^{\alpha^\prime} \\ g_i^{{\tilde{\alpha}}^\prime}
\end{array} \right)
=\epsilon_i
\left( \begin{array}{c} f_i^{\alpha^\prime} \\ g_i^{{\tilde{\alpha}}^\prime}
\end{array} \right)\;,
\end{equation}
of dimension: $\alpha_{max}+\tilde{\alpha}_{max}$.

The time-reversal operator $\hat{T}=-i\sigma_y \hat{K}$
exchanges  the simplex eigenvalues:
\begin{equation}
\hat{T}\Phi_{\alpha}(\mathbf{r};+)=-\Phi_{\alpha}(\mathbf{r};-)
\quad \textnormal{and} \quad
\hat{T}\Phi_{\alpha}(\mathbf{r};-)=\Phi_{\alpha}(\mathbf{r};+) \;,
\end{equation}
and, thus, when acting on the Dirac spinors
\begin{equation}
\hat{T}\psi_{i}(\mathbf{r};+)=
 \hat{T}  \left( \begin{array}{c}f_{i}(\mathbf{r};+) \\
                                ig_{i}(\mathbf{r};-) \end{array} \right)
     =   \left( \begin{array}{c}\hat{T} f_i(\mathbf{r};+) \\
                             -i \hat{T}g_i(\mathbf{r};-) \end{array} \right)
           =    \left( \begin{array}{c} -f_i(\mathbf{r};-) \\
                                        -ig_i(\mathbf{r};+) \end{array} \right)
           =- \psi_i(\mathbf{r};-) \; .
\end{equation}
Time-reversed single particle states correspond to opposite simplex
eigenvalues. Because of time-reversal invariance, for each solution
of the Dirac equation (\ref{dirac-app}) with positive simplex
eigenvalue $\psi_i(\mathbf{r};+)$, there exists a degenerate
time-reversed solution with negative simplex eigenvalue
$\psi_i(\mathbf{r};-)$. Both solutions  contribute equally to the
densities, and in practice only the Dirac equation for positive
simplex eigenstates is solved.

In the current implementation of the model, parity is also imposed as
a conserved symmetry. This means that the basis states split into two
parity blocks, which can be diagonalized separately. In addition, we
require that the densities are symmetric under reflections with
respect to the $yz$, $xz$ and $xy$ planes.
\begin{equation}
\label{symmetries-rho}
\rho_{s,v}(x,y,z)=\rho_{s,v}(-x,y,z)=\rho_{s,v}(x,-y,z)=\rho_{s,v}(x,y,-z)\; .
\end{equation}
The symmetries of the scalar and vector densities are, of course,
fulfilled by the corresponding self-consistent scalar and vector potentials:
\begin{align}
\label{symmetries-sca-pot}
S(x,y,z)&=S(-x,y,z)=S(x,-y,z)=S(x,y,-z) \\
\label{symmetries-vec-pot}
V(x,y,z)&=V(-x,y,z)=V(x,-y,z)=V(x,y,-z)\;.
\end{align}
The self-consistent symmetries (\ref{symmetries-sca-pot}) and
(\ref{symmetries-vec-pot}) simplify the evaluation of the matrix
elements $\mathcal{A}_{\alpha\alpha^\prime}$ and
$\mathcal{C}_{\tilde{\alpha}\tilde{\alpha}^\prime}$. First, the
symmetry under reflections with respect to the $xy$, $yz$ and $xz$
planes means that we need to calculate only matrix elements between
states $\phi_{\alpha}$ and $\phi_{\alpha^\prime}$ for which:
\begin{equation}
(-1)^{n_x}=(-1)^{n_x^\prime}\;, \quad (-1)^{n_y}=(-1)^{n_y^\prime} \quad
\textnormal{and} \quad (-1)^{n_z}=(-1)^{n_z^\prime}\;.
\label{parities}
\end{equation}
Furthermore, three-dimensional integrals reduce to the octant $x,y,z\ge0$.
The matrix elements of the vector and scalar potentials read:
\begin{align}
\label{matA}
\mathcal{A}_{\alpha \alpha^\prime} &=
\left\langle \alpha;+\right| m^* + V\left|\alpha^\prime;+\right\rangle\\
&= 8(-1)^{(n_y-n_y^\prime)/2}\int_0^\infty{\int_0^\infty{\int_0^\infty{
\phi_{n_x}(x)\phi_{n_y}(y)\phi_{n_z}(z) (V+m^*)
\phi_{n_x^\prime}(x)\phi_{n_y^\prime}(y)\phi_{n_z^\prime}(z)dV}} }
        \nonumber \\
\label{matC}
\mathcal{C}_{\tilde{\alpha} \tilde{\alpha}^\prime} &=
 \left\langle \tilde{\alpha};-\left| V-m^* \right|
 \tilde{\alpha}^\prime;-\right\rangle \\
 &= 8(-1)^{(n_y-n_y^\prime)/2}\int_0^\infty{\int_0^\infty{\int_0^\infty{
   \phi_{n_x}(x)\phi_{n_y}(y)\phi_{n_z}(z) (V-m^*)
 \phi_{n_x^\prime}(x)\phi_{n_y^\prime}(y)\phi_{n_z^\prime}(z)dV}} } \;.
            \nonumber
\end{align}
Note that the condition (\ref{parities}) means that the difference
$(n_y-n_y^\prime)$ must be
even, hence the matrix elements (\ref{matA}) and (\ref{matC}) are real.

The matrix elements of the kinetic energy term
$\mathcal{B}_{\alpha\tilde{\alpha}^\prime}$ can
be calculated analytically using the expression:
\begin{equation}
\partial_\mu \phi_{n_\mu}(x_\mu) = \frac{1}{\sqrt{2}b_\mu}
    \left[-\sqrt{n_\mu+1}\phi_{n_\mu+1}(x_\mu)
          +\sqrt{n_\mu}\phi_{n_\mu-1}(x_\mu)   \right]\;,
\end{equation}
together with the orthogonality relation (\ref{orthogonality}). The operator
$\mathbf{\sigma} \cdot \mathbf{\nabla}$ consists of three terms:
\begin{equation}
\mathcal{B}_{\alpha \tilde{\alpha}^\prime}
  =  \left\langle \alpha; +\left|
  -\sigma_x\partial_x - \sigma_y\partial_y-\sigma_z\partial_z
     \right| \tilde{\alpha}^\prime;-\right\rangle
 =   \mathcal{B}^x_{\alpha \tilde{\alpha}^\prime}
 +  \mathcal{B}^y_{\alpha \tilde{\alpha}^\prime}
 +  \mathcal{B}^z_{\alpha \tilde{\alpha}^\prime} \; ,
\end{equation}
and the corresponding matrix elements are calculated from
\begin{align}
\mathcal{B}^x_{\alpha \tilde{\alpha}^\prime} &=(-1)^{n_y}
\delta_{n_y n_y^\prime}\delta_{n_z n_z^\prime}
\frac{1}{\sqrt{2}b_x}\left[-\sqrt{n_x^\prime+1}\delta_{n_x n_x^\prime+1}
       +\sqrt{n_x^\prime}\delta_{n_x n_x^\prime-1}  \right],
\\
 \mathcal{B}^y_{\alpha \tilde{\alpha}^\prime}&=(-1)^{n_y^\prime}
 \delta_{n_x n_x^\prime}\delta_{n_z n_z^\prime}
\frac{1}{\sqrt{2}b_y}\left[\sqrt{n_y^\prime+1}\delta_{n_y n_y^\prime+1}
       +\sqrt{n_y^\prime}\delta_{n_y n_y^\prime-1}  \right],
 \\
 \mathcal{B}^z_{\alpha \tilde{\alpha}^\prime}&=(-1)^{n_x^\prime+n_y^\prime+1}
   \delta_{n_x n_x^\prime}\delta_{n_y n_y^\prime}
 \frac{1}{\sqrt{2}b_z}\left[-\sqrt{n_z^\prime+1}\delta_{n_z n_z^\prime+1}
       +\sqrt{n_z^\prime}\delta_{n_z n_z^\prime-1}  \right] \; .
\end{align}

The set of self-consistent solutions of the single-nucleon Dirac equation
determines the scalar and vector densities:
\begin{align}
\rho_v(\mathbf{r}) &=
   2\sum_{i}{v_i^2 \psi_i^\dagger(\mathbf{r};+)\psi_i(\mathbf{r};+)},\\
\rho_s(\mathbf{r}) &=
   2\sum_{i}{v_i^2 \psi_i^\dagger(\mathbf{r};+)\beta\psi_i(\mathbf{r};+)}.
\end{align}
Because of time-reversal symmetry, the summation is over positive
simplex solutions. To calculate densities in coordinate space, one
needs explicit expressions for the products of basis states:
\begin{align}
\label{product-plus}
\phi^\dagger_\alpha(\mathbf{r};+)\phi_{\alpha^\prime}(\mathbf{r};+)
 &= (-1)^{(n_y^\prime-n_y)/2}\phi_{n_x}(x)\phi_{n_x^\prime}(x)
     \phi_{n_y}(y)\phi_{n_y^\prime}(y)\phi_{n_z}(z)\phi_{n_z^\prime}(z),\\
\label{product-minus}
\phi^\dagger_\alpha(\mathbf{r};-)\phi_{\alpha^\prime}(\mathbf{r};-)
 &= (-1)^{(n_y-n_y^\prime)/2}\phi_{n_x}(x)\phi_{n_x^\prime}(x)
     \phi_{n_y}(y)\phi_{n_y^\prime}(y)\phi_{n_z}(z)\phi_{n_z^\prime}(z).
\end{align}
Note that the symmetry requirement Eq.~(\ref{symmetries-rho}) imposes
the condition (\ref{parities}) and, again, the difference
$(n_y-n_y^\prime)$ is even so that the contributions
(\ref{product-plus}) and (\ref{product-minus})  to the densities are
both real.

\section{\label{app-B}Moments of inertia}

The basic ingredient of the Inglis-Belyaev formula for the moments of
inertia Eq.~(\ref{Inglis-Belyaev}) are the matrix elements of the
angular momentum operators in the simplex basis (\ref{basis-pos}) and
(\ref{basis-neg}). Here we present in detail the calculation of the
matrix element of the $\hat{J}_x$ component between basis states with
positive simplex eigenvalue
\begin{equation}
\langle \alpha;+|\hat{J}_x|\alpha^\prime;+\rangle \; .
\end{equation}
For the other matrix elements only the final expressions will be listed.

The $x$ component of the total angular momentum operator is the sum of
the spin and the spatial contributions:
\begin{equation}
\hat{J}_x =  \frac{\hbar}{2}\hat{\sigma}_x + \hat{L}_x =
               \frac{\hbar}{2}\hat{\sigma}_x
             -i\hbar\left(y\partial_z-z\partial_y  \right)\;.
\end{equation}
The spatial parts of the basis states are unaffected by the $\hat{\sigma}_x$
operator, thus generating the product of Kronecker delta symbols
$\delta_{n_x,n_x^\prime}\delta_{n_y,n_y^\prime}\delta_{n_z,n_z^\prime}$.
The contribution from the spin factors of the basis states is given by
\begin{equation}
[ \chi^\dagger_+ - (-1)^{n_x}\chi^\dagger_- ] \hat{\sigma}_x
[ \chi_+ - (-1)^{n_x}\chi_-]=(-1)^{n_x+1} + (-1)^{n_x^\prime+1} \; ,
\end{equation}
and the spin matrix element reads:
\begin{equation}
\frac{\hbar}{2}
  \langle \alpha;+| \hat{\sigma}_x |\alpha^\prime;+ \rangle
= \frac{\hbar}{2}
 \delta_{n_x,n_x^\prime}\delta_{n_y,n_y^\prime}\delta_{n_z,n_z^\prime}
         (-1)^{n_x+1} \; .
\end{equation}
Next, the contribution from the operator $\hat{L}_x$ is calculated
\begin{equation}
\label{mat-Lx}
\langle \alpha;+| \hat{L}_x | \alpha^\prime;+ \rangle
= -i\hbar
\langle \alpha;+|y\partial_z-z\partial_y  | \alpha^\prime;+ \rangle.
\end{equation}
The spin factors of the basis states are not affected by the $\hat{L}_x$ operator:
\begin{equation}
[ \chi^\dagger_+ - (-1)^{n_x}\chi^\dagger_- ]
[ \chi_+ - (-1)^{n_x}\chi_-]= 1+(-1)^{n_x+n_x^\prime}.
\end{equation}
To calculate the matrix elements Eq.~(\ref{mat-Lx}), the following relations are used:
\begin{align}
x_\mu \phi_{n_\mu} &=\frac{b_\mu}{\sqrt{2}}\left[ \sqrt{n_\mu+1}\phi_{n_\mu+1}(x_\mu)
        +\sqrt{n_\mu}\phi_{n_\mu-1}(x_\mu)   \right], \\
\partial_\mu \phi_{n_\mu}(x_\mu) &= \frac{1}{\sqrt{2}b_\mu}
    \left[-\sqrt{n_\mu+1}\phi_{n_\mu+1}(x_\mu)
          +\sqrt{n_\mu}\phi_{n_\mu-1}(x_\mu)   \right],
\end{align}
together with the orthonormality relation (\ref{orthogonality}).
The total matrix element reads
\begin{align}
\langle \alpha;+|\hat{J}_x |\alpha^\prime;+\rangle
&=  \frac{\hbar}{2}(-1)^{n_x+1}
      \delta_{n_x,n_x^\prime}\delta_{n_y,n_y^\prime}
        \delta_{n_z,n_z^\prime} \\
&+ \frac{\hbar}{2}  \delta_{n_x,n_x^\prime}\left(\frac{b_y}{b_z} -\frac{b_z}{b_y} \right)
\left[ \sqrt{n_y^\prime+1}\sqrt{n_z^\prime+1}
        \delta_{n_y,n_y^\prime+1}\delta_{n_z,n_z^\prime+1}
      +\sqrt{n_y^\prime}\sqrt{n_z^\prime}
        \delta_{n_y,n_y^\prime-1}\delta_{n_z,n_z^\prime-1}\right]
          \nonumber \\
&- \frac{\hbar}{2}  \delta_{n_x,n_x^\prime}\left(\frac{b_y}{b_z} +\frac{b_z}{b_y} \right)
\left[ \sqrt{n_y^\prime+1}\sqrt{n_z^\prime}
        \delta_{n_y,n_y^\prime+1}\delta_{n_z,n_z^\prime-1}
      +\sqrt{n_y^\prime}\sqrt{n_z^\prime+1}
        \delta_{n_y,n_y^\prime-1}\delta_{n_z,n_z^\prime+1}\right] \; .
 \nonumber
\end{align}
The following relations can easily be proved:
\begin{equation}
\langle\alpha;+|\hat{J}_x |\alpha^\prime;+\rangle
=-\langle\alpha;-|\hat{J}_x |\alpha^\prime;-\rangle,\quad
\textnormal{and} \quad
\langle \alpha;+|\hat{J}_x |\alpha^\prime;-\rangle =
\langle \alpha;-|\hat{J}_x |\alpha^\prime;+\rangle =0 \; .
\end{equation}
The final expression for the moment of inertia
$\mathcal{I}_x\equiv \mathcal{I}_1$:
\begin{equation}
\mathcal{I}_x =2\sum_{i,j>0}{\frac{\left(u_iv_j-v_iu_j \right)^2}{E_i+E_j}
\left| \sum_{\alpha \alpha^\prime}f_i^\alpha f_j^{\alpha^\prime}
 \langle \alpha;+|\hat{J}_x |\alpha^\prime;+\rangle
- \sum_{\tilde{\alpha} \tilde{\alpha}^\prime}
   g_i^{\tilde{\alpha}} g_j^{\tilde{\alpha}^\prime}
 \langle \tilde{\alpha};+|\hat{J}_x |\tilde{\alpha}^\prime;+\rangle
 \right|^2}.
\end{equation}
We include only the final results for the matrix elements of $\hat{J}_y$ and $\hat{J}_z$:
\begin{align}
\langle \alpha;+|\hat{J}_y |\alpha^\prime;-\rangle
&=  i\frac{\hbar}{2}(-1)^{n_y}
      \delta_{n_x,n_x^\prime}\delta_{n_y,n_y^\prime}
      \delta_{n_z,n_z^\prime} \\
&+ i\frac{\hbar}{2}  \delta_{n_y,n_y^\prime}(-1)^{n_x+n_y}
     \left(\frac{b_z}{b_x} -\frac{b_x}{b_z} \right) \times \nonumber \\
   &\times   \left[ \sqrt{n_x^\prime+1}\sqrt{n_z^\prime+1}
        \delta_{n_x,n_x^\prime+1}\delta_{n_z,n_z^\prime+1}
      -\sqrt{n_y^\prime}\sqrt{n_z^\prime}
        \delta_{n_x,n_x^\prime-1}\delta_{n_z,n_z^\prime-1} \right]
     \nonumber \\
&+i\frac{\hbar}{2}  \delta_{n_y,n_y^\prime}(-1)^{n_x+n_y}
    \left(\frac{b_z}{b_x} +\frac{b_x}{b_z} \right)\times \nonumber \\
  &\times   \left[ \sqrt{n_x^\prime+1}\sqrt{n_z^\prime}
        \delta_{n_x,n_x^\prime+1}\delta_{n_z,n_z^\prime-1}
      -\sqrt{n_x^\prime}\sqrt{n_z^\prime+1}
        \delta_{n_x,n_x^\prime-1}\delta_{n_z,n_z^\prime+1} \right],
\nonumber
\end{align}
\begin{equation}
\langle \alpha;-|\hat{J}_y |\alpha^\prime;+\rangle
=-\langle \alpha;+|\hat{J}_y |\alpha^\prime;-\rangle,\quad
\textnormal{and} \quad
\langle \alpha;+|\hat{J}_y |\alpha^\prime;+\rangle =
\langle \alpha;-|\hat{J}_y |\alpha^\prime;-\rangle =0,
\end{equation}
\begin{align}
\langle \alpha;+|\hat{J}_z |\alpha^\prime;-\rangle
&=  \frac{\hbar}{2}(-1)^{n_x+n_y+1}
      \delta_{n_x,n_x^\prime}\delta_{n_y,n_y^\prime}
      \delta_{n_z,n_z^\prime} \\
&+ \frac{\hbar}{2}  \delta_{n_z,n_z^\prime}(-1)^{n_x+n_y+1}
     \left(\frac{b_x}{b_y} -\frac{b_y}{b_x} \right) \times \nonumber \\
   &\times   \left[ \sqrt{n_x^\prime+1}\sqrt{n_y^\prime+1}
        \delta_{n_x,n_x^\prime+1}\delta_{n_y,n_y^\prime+1}
      +\sqrt{n_x^\prime}\sqrt{n_y^\prime}
        \delta_{n_x,n_x^\prime-1}\delta_{n_y,n_y^\prime-1} \right]
      \nonumber \\
&+ \frac{\hbar}{2}  \delta_{n_z,n_z^\prime}(-1)^{n_x+n_y+1}
    \left(\frac{b_x}{b_y} +\frac{b_y}{b_x} \right)\times \nonumber \\
  &\times   \left[ \sqrt{n_x^\prime+1}\sqrt{n_y^\prime}
        \delta_{n_x,n_x^\prime+1}\delta_{n_y,n_y^\prime-1}
      +\sqrt{n_x^\prime}\sqrt{n_y^\prime+1}
        \delta_{n_x,n_x^\prime-1}\delta_{n_y,n_y^\prime+1} \right],
 \nonumber
\end{align}
\begin{equation}
\langle \alpha;-|\hat{J}_z | \alpha^\prime;+\rangle
=\langle \alpha;+|\hat{J}_z |\alpha^\prime;-\rangle,\quad
\textnormal{and} \quad
\langle \alpha;+|\hat{J}_z | \alpha^\prime;+\rangle =
\langle \alpha;-|\hat{J}_z |\alpha^\prime;-\rangle =0,
\end{equation}
The corresponding moments of inertia $\mathcal{I}_y \equiv \mathcal{I}_2$ and
$\mathcal{I}_z \equiv \mathcal{I}_3$ read:
\begin{align}
\mathcal{I}_y &= 2\sum_{i,j>0}{\frac{\left(u_iv_j-v_iu_j\right)^2}{E_i+E_j}
\left| \sum_{\alpha \alpha^\prime}f_i^\alpha f_j^{\alpha^\prime}
 \langle \alpha;+|\hat{J}_y | \alpha^\prime;-\rangle
- \sum_{\tilde{\alpha} \tilde{\alpha}^\prime}g_i^{\tilde{\alpha}} g_j^{\tilde{\alpha}^\prime}
\langle \tilde{\alpha};+|\hat{J}_y |\tilde{\alpha}^\prime;-\rangle
 \right|^2}, \\
\mathcal{I}_z &=2\sum_{i,j>0}{\frac{\left(u_iv_j-v_iu_j\right)^2}{E_i+E_j}
\left| \sum_{\alpha \alpha^\prime}f_i^\alpha f_j^{\alpha^\prime}
 \langle \alpha;+|\hat{J}_z |\alpha^\prime;-\rangle
+ \sum_{\tilde{\alpha} \tilde{\alpha}^\prime}g_i^{\tilde{\alpha}} g_j^{\tilde{\alpha}^\prime}
 \langle \tilde{\alpha};+|\hat{J}_z |\tilde{\alpha}^\prime;-\rangle
 \right|^2}.
\end{align}
All three moments of inertia vanish at the spherical point $\beta=0$.
In addition, $\mathcal{I}_z$ vanishes for the $\gamma=0^0$
configurations (prolate deformed, with $z$ as the symmetry axis),
whereas $\mathcal{I}_y$ vanishes at $\gamma = 60^0$ (oblate deformed,
$y$ is the symmetry axis). These conditions are incorporated in the
following functional form:
\begin{equation}
\mathcal{I}_k = 4B_k\beta^2 \sin^2{(\gamma-2k\pi/3)}\;,  \textnormal{ with }
(1\equiv x,\;2\equiv y,\;3\equiv z) \; ,
\end{equation}
from which the inertia parameters $B_x$, $B_y$ and $B_z$ follow:
\begin{equation}
\label{Bx}
B_k = \frac{\mathcal{I}_k}{4\beta^2 \sin^2{(\gamma-2k\pi/3)}}.
\end{equation}
For the limiting cases described above the following relations are used
\begin{equation}
B_x(\beta=0)=B_y(\beta=0)=B_z(\beta=0)=B_{\gamma\gamma}(\beta=0),
\end{equation}
\begin{equation}
B_y(\beta,\gamma=60^0)=B_{\gamma\gamma}(\beta,\gamma=60^0),
\end{equation}
\begin{equation}
B_z(\beta,\gamma=0^0)=B_{\gamma\gamma}(\beta,\gamma=0^0).
\end{equation}

To calculate the mass parameters from Eqs. (\ref{masspar-B}) and
(\ref{masspar-M}), one needs the matrix elements of the operators
$x^2$, $y^2$ and $z^2$ in the simplex basis. These are combined 
and inserted into the matrix elements
\begin{equation}
\left<\psi_i\right| \hat{Q}_{2\mu} \left| \psi_j \right> =
\sum_{\alpha \alpha^\prime}f_i^\alpha f_j^{\alpha^\prime}
 \langle \alpha;+|\hat{Q}_{2\mu} | \alpha^\prime;+\rangle
+ \sum_{\tilde{\alpha} \tilde{\alpha}^\prime}g_i^{\tilde{\alpha}} g_j^{\tilde{\alpha}^\prime}
\langle \tilde{\alpha};+|\hat{Q}_{2\mu} |\tilde{\alpha}^\prime;+\rangle \;,
\end{equation}
that determine the $2\times 2$ matrix $\mathcal{M}_{(n),\mu\nu}$
Eq.~(\ref{masspar-M}). The mass parameters $B_{\mu\nu}(q_0,q_2)$
are then calculated from Eq.~(\ref{masspar-B}), and
transformed from the quadrupole coordinates $q_0,q_2$ to the
polar Bohr deformation variables $\beta$ and $\gamma$.
\section{\label{app-E}Rotational zero-point energy correction}

The rotational zero-point energy Eq.~(\ref{ZPE-rot}) is determined by
the matrix elements of the quadrupole operators:
\begin{equation}
\hat{Q}_{21}=-2iyz\; , \quad \hat{Q}_{2-1}=-2xz\; , \quad \textnormal{and} \quad
\hat{Q}_{2-2}=2ixy \;.
\end{equation}
Using the following expression:
\begin{equation}
x_\mu \phi_{n_\mu}(x_\mu) =
\frac{b_\mu}{\sqrt{2}}\left[ \sqrt{n_\mu+1}\phi_{n_\mu+1}(x_\mu)
        +\sqrt{n_\mu}\phi_{n_\mu-1}(x_\mu)   \right],
\end{equation}
together with the orthogonality relation (\ref{orthogonality}),
the calculation of matrix elements is straightforward. Here
we list only the final expressions:
\begin{align}
\langle\alpha;+|\hat{Q}_{21}|\alpha^\prime;+\rangle
 &= b_yb_z\delta_{n_x,n_x^\prime}
     \left[-\sqrt{n_y^\prime+1}\delta_{n_y,n_y^\prime+1}
           +\sqrt{n_y^\prime} \delta_{n_y,n_y^\prime-1}\right] \times \nonumber\\
    &\times  \left[\sqrt{n_z^\prime+1}\delta_{n_z,n_z^\prime+1}
           +\sqrt{n_z^\prime} \delta_{n_z,n_z^\prime-1}\right]     \\
\langle\alpha;-|\hat{Q}_{21}|\alpha^\prime;-\rangle
&=-\langle\alpha;+|\hat{Q}_{21}|\alpha^\prime;+\rangle  \\
\langle\alpha;+|\hat{Q}_{21}|\alpha^\prime;-\rangle&=
\langle\alpha;-|\hat{Q}_{21}|\alpha^\prime;+\rangle=0 \\
\langle\alpha;+|\hat{Q}_{2-1}|\alpha^\prime;-\rangle
& = (-1)^{n_x^\prime+n_y^\prime}b_xb_z\delta_{n_y,n_y^\prime}
     \left[\sqrt{n_x^\prime+1}\delta_{n_x,n_x^\prime+1}
           +\sqrt{n_x^\prime} \delta_{n_x,n_x^\prime-1}\right]\times \nonumber\\
    &\times   \left[\sqrt{n_z^\prime+1}\delta_{n_z,n_z^\prime+1}
           +\sqrt{n_z^\prime} \delta_{n_z,n_z^\prime-1}\right]\\
\langle\alpha;-|\hat{Q}_{2-1}|\alpha^\prime;+\rangle
&=-\langle\alpha;+|\hat{Q}_{2-1}|\alpha^\prime;-\rangle \\
\langle\alpha;+|\hat{Q}_{2-1}|\alpha^\prime;+\rangle&=
\langle\alpha;-|\hat{Q}_{2-1}|\alpha^\prime;-\rangle=0\\
\langle\alpha;+|\hat{Q}_{2-2}|\alpha^\prime;-\rangle
 &= (-1)^{n_x^\prime+n_y^\prime+1}b_xb_y\delta_{n_z,n_z^\prime}
     \left[\sqrt{n_x^\prime+1}\delta_{n_x,n_x^\prime+1}
           +\sqrt{n_x^\prime} \delta_{n_x,n_x^\prime-1}\right]\times \nonumber\\
   &\times    \left[\sqrt{n_y^\prime+1}\delta_{n_y,n_y^\prime+1}
           -\sqrt{n_y^\prime} \delta_{n_y,n_y^\prime-1}\right]  \\
\langle\alpha;-|\hat{Q}_{2-2}|\alpha^\prime;-\rangle
&=-\langle\alpha;+|\hat{Q}_{2-2}|\alpha^\prime;+\rangle   \\
\langle \alpha;+|\hat{Q}_{2-2}|\alpha^\prime;-\rangle&=
\langle\alpha;-|\hat{Q}_{2-2}|\alpha^\prime;+\rangle=0
\end{align}
\section{\label{app-F}Numerical details}

\begin{figure}
\includegraphics[scale=0.6,angle=270]{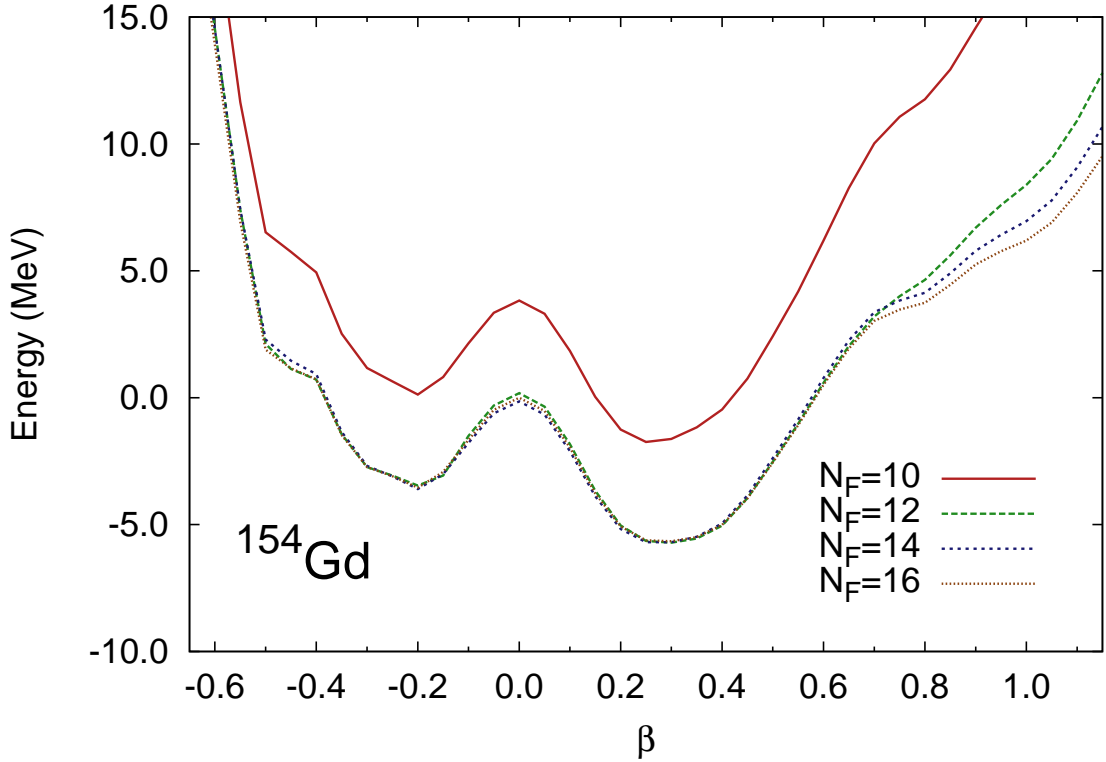}
\vspace{2cm}
\caption{(Color online)
Self-consistent RMF+BCS binding energy curves of $^{154}$Gd,
as functions of the axial deformation parameter $\beta$. Negative values
of $\beta$ correspond to the $(\beta>0,\;\gamma=180^0)$ axis on the
$\beta - \gamma$ plane. The four energy curves correspond to
calculations in the three-dimensional harmonic oscillator basis with
$10$, $12$, $14$ and $16$ major oscillator shells, respectively.}
\label{Fig23}
\end{figure}

In Fig.~\ref{Fig23} we check the convergence of the RMF+BCS quadrupole
constrained calculations as a function of the maximal number of
oscillator shells used in the expansion of the Dirac spinors. The
binding energy curves for $^{154}$Gd are plotted as functions of the
axial $\beta$ deformation. These curves correspond to calculations
with $10, 12$, $14$ and $16$ major oscillator shells. For moderate
deformations considered
 in this study ($|\beta|<0.65$), the RMF results show convergence at $N_{F}=14$.

The self-consistent RMF+BCS equations are solved on a mesh over
the first sextant of the $\beta - \gamma$ plane:
\begin{equation}
\beta \ge 0;\;\Delta \beta =0.05;\;  0 \le \gamma \le 60^0;\;
\Delta \gamma=6^0 \; ,
\label{equidistant-mesh}
\end{equation}
and the maximum value of the $\beta$ deformation is:
$\beta_{max}=1.15$.

The choice of the basis wave functions Eq.~(\ref{basis-start})
introduces a factor $e^{-\mu^2\beta^2}$ into the integral over
$\beta$ in the matrix elements of the collective Hamiltonian. With
the substitution $y\equiv \mu^2\beta^2$, we obtain the weight
function appropriate for Gauss-Laguerre quadrature. The integrals
over $\gamma$ are evaluated by Gauss-Legendre quadrature. The
corresponding number of mesh points are: $n_\beta=24$ and
$n_\gamma=24$, respectively. The parameters of the collective
Hamiltonian in the Gaussian mesh-points are calculated by
interpolation from the values calculated on the equidistant mesh
Eq.~(\ref{equidistant-mesh}). To avoid extrapolation, the minimum
value of the basis parameter $\mu$ is restricted to:
\begin{equation}
\frac{\sqrt{y_{n_\beta}}}{\mu} < \beta_{max} \; .
\end{equation}
We have verified that both the calculated excitation spectra and
transition probabilities remain stable for any  choice of $\mu$ within the
interval: $8 \le\mu\le 15$. All calculations presented in this work
are performed with $\mu=9$.

\bigskip \bigskip
\leftline{\bf ACKNOWLEDGMENTS}
We thank Jean Libert for useful discussions.
This work was supported in part by MZOS - project 1191005-1010,
by the Major State 973 Program 2007CB815000,
the NSFC under Grant Nos. 10435010, 10775004 and 10221003,
and by the DFG cluster of excellence \textquotedblleft Origin and
Structure of the Universe\textquotedblright\ (www.universe-cluster.de).
Z. P. Li acknowledges support from the Croatian National Foundation 
for Science. The work of J.M, T.N., and D.V. was supported in part by 
the Chinese-Croatian project "Nuclear structure far from stability". 
\bigskip


\clearpage
\clearpage
\clearpage
\clearpage
\clearpage
\clearpage
\clearpage
\clearpage
\clearpage
\clearpage
\clearpage
\clearpage
\clearpage
\clearpage
\clearpage
\clearpage
\clearpage
\clearpage
\clearpage
\clearpage
\clearpage
\clearpage
\clearpage
\clearpage
\clearpage
\begin {table}[tbp]
\begin {center}
\caption{Average values of the deformation parameters $\beta$ and 
$\gamma$ (cf. Eqs.~(\ref{avbeta}) and (\ref{avgamma})) for the 
calculated first, second, and third $2^+$, $4^+$ and $6^+$ states in 
$^{152-160}$Gd.}
\bigskip
\begin {tabular}{c|c|cc|cc|cc}
\hline
\multicolumn{2}{c}{} &
\multicolumn{2}{c}{ $ 2^+$} &
\multicolumn{2}{c}{ $ 4^+ $} & 
\multicolumn{2}{c}{$  6^+$ }
  \\ \hline
 & state   & $\langle \beta\rangle $ & $\langle \gamma\rangle $ (deg)
                & $\langle \beta\rangle $ & $\langle \gamma\rangle $ (deg)
                & $\langle \beta\rangle $ & $\langle \gamma\rangle $ (deg)\\
\hline 
\multirow{3}{*}{$^{152}$Gd}& $J_1^+$ & 0.24 & 17.0 & 0.26 & 15.3
                                                               & 0.28 & 13.7\\
                                            & $J_2^+$ & 0.26 & 19.0 & 0.31 & 13.9 
                                                               & 0.32 & 12.8\\
                                            & $J_3^+$ & 0.29 & 16.3& 0.28 & 20.5 
                                                               & 0.30 & 20.1\\ \hline
\multirow{3}{*}{$^{154}$Gd}& $J_1^+$ & 0.31 & 12.8 & 0.32 & 12.0 
                                                               & 0.33 & 11.3\\
                                            & $J_2^+$ & 0.33 & 12.8 & 0.33 & 12.4
                                                               & 0.34 & 12.2\\
                                            & $J_3^+$ & 0.29 & 19.8 & 0.32 & 17.6
                                                               & 0.34 & 16.2\\ \hline    
\multirow{3}{*}{$^{156}$Gd}& $J_1^+$ & 0.34 & 11.3 & 0.35 & 11.0
                                                               & 0.36 & 10.6\\
                                            & $J_2^+$ & 0.34 & 13.0 & 0.34 & 14.0 
                                                               & 0.35 & 13.6\\
                                            & $J_3^+$ & 0.35 & 13.0 & 0.36 & 12.5 
                                                               & 0.37 & 12.0\\ \hline   
\multirow{3}{*}{$^{158}$Gd}& $J_1^+$ & 0.36 & 10.8 & 0.36 & 10.6 
                                                               & 0.36 & 10.5\\
                                            & $J_2^+$ & 0.36 & 14.3 & 0.36 & 13.8
                                                               & 0.37 & 13.4\\
                                            & $J_3^+$ & 0.36 & 11.0 & 0.37 & 10.7 
                                                               & 0.38 & 10.4\\ \hline 
\multirow{3}{*}{$^{160}$Gd}& $J_1^+$ & 0.36 & 10.3 & 0.37 & 10.2 
                                                               & 0.37 & 10.1\\
                                            & $J_2^+$ & 0.37 & 13.4 & 0.37 & 13.2 
                                                               & 0.37 & 12.9\\
                                            & $J_3^+$ & 0.38 & 10.3 & 0.38 & 10.0 
                                                               & 0.39 & 9.8\\                                                                                                                                                                                                                                                                                                    
\end{tabular}
\label{Tab-B}
\end{center}
\end{table}
\end{document}